\documentclass[prd,aps,preprintnumbers,nofootinbib,superscriptaddress]{revtex4}
\pdfoutput=1
\usepackage{amssymb,amsmath,graphicx,bm}

\DeclareMathOperator{\tr}{Tr}

\usepackage[usenames,dvipsnames]{color}
\usepackage[normalem]{ulem} 
\renewcommand\sout{\bgroup \color[rgb]{0.55,0.00,0.99} \ULdepth=-.5ex \ULset}
\newcommand{\xB}{x_{\scriptscriptstyle B}}

\newcommand{\sT}{{\scriptscriptstyle T}}
\renewcommand{\d}{\mathrm{d}}
\def\slash#1{\setbox0=\hbox{$#1$}               
        \dimen0=\wd0                            
        \setbox1=\hbox{/} \dimen1=\wd1          
        \ifdim\dimen0>\dimen1                   
        \rlap{\hbox to \dimen0{\hfil/\hfil}}    
        #1                                      
        \else
        \rlap{\hbox to \dimen1{\hfil$#1$\hfil}} 
        /                                       
        \fi}                                    %

\newcommand{\qs}{q \!\!\! /}
 
\newcommand{\ps}{p \!\!\! /}
\newcommand{\Pqs}{{P \!\!\! \!/}_{\cal Q}}
\newcommand{\epss}{\varepsilon\!\!\!/}

\usepackage[normalem]{ulem} 
\renewcommand\sout{\bgroup \color[rgb]{0.55,0.00,0.99} \ULdepth=-.5ex \ULset}

\begin{document}

\title{Gluon TMDs and NRQCD matrix elements in $J/\psi$ production at  an EIC}

\author{Alessandro~Bacchetta}
\email{alessandro.bacchetta@unipv.it}
\affiliation{Dipartimento di Fisica, Universit\`a di Pavia,  via Bassi 6, I-27100 Pavia, Italy} 
\affiliation{INFN Sezione di Pavia, via Bassi 6, I-27100 Pavia, Italy}

\author{Dani\"el Boer}
\email{d.boer@rug.nl}
\affiliation{ {Van Swinderen Institute for Particle Physics and Gravity, University of Groningen, Nijenborgh 4, 9747 AG Groningen, The Netherlands}}

\author{Cristian Pisano}
\email{cristian.pisano@ca.infn.it}
\affiliation{Dipartimento di Fisica, Universit\`a di Cagliari, Cittadella Universitaria, I-09042 Monserrato (CA), Italy}
\affiliation{INFN Sezione di Cagliari,  Cittadella Universitaria, I-09042 Monserrato (CA), Italy}

\author{Pieter Taels}
\email{pieter.taels@pv.infn.it}
\affiliation{INFN Sezione di Pavia, via Bassi 6, I-27100 Pavia, Italy}

\begin{abstract}
In this paper we analyze azimuthal asymmetries in the processes of unpolarized and polarized $J/\psi \,(\Upsilon)$ production at an Electron-Ion Collider. Apart from giving access to various unknown gluon transverse momentum distributions, we suggest to use them as a new method to extract specific color-octet NRQCD long-distance matrix elements, i.e.\ $\langle0\vert{\cal O}_{8}^{J/\psi}(^{1}S_{0})\vert0\rangle$ and $\langle0\vert{\cal O}_{8}^{J/\psi}(^{3}P_{0})\vert0\rangle$, whose values are still quite uncertain and for which lattice calculations are unavailable. The new method is based on combining measurements of analogous asymmetries in open heavy-quark pair production which can be performed at the same energy. We also study for the first time the effects of transverse-momentum smearing in the quarkonium formation process. To enhance the gluon contribution one can consider smaller values of $x$ and, in order to assess the impact of small-$x$ evolution, we perform a numerical study using the MV model as a starting input and evolve it with the JIMWLK equations.
\end{abstract}

\date{\today}

\maketitle

\section{Introduction}

Transverse momentum dependent parton distributions (TMDs) are fundamental objects which encode information on the motion of partons inside hadrons and on the correlations between spin and partonic transverse momenta. As such, they can be considered as an extension of the standard, one-dimensional, parton distribution functions (PDFs) to the three-dimensional momentum space. Contrary to PDFs, TMDs are in general not universal. This is due to their sensitivity to the soft gluon exchanges and the color flow in the specific process in which they are probed. A typical example is provided by the Sivers function for quarks~\cite{Sivers:1989cc}, namely the azimuthal distribution of unpolarized quarks inside a transversely polarized proton, which is expected to enter with opposite sign in the single spin asymmetries for semi-inclusive deep inelastic scattering (SIDIS) and for the Drell-Yan processes~\cite{Brodsky:2002rv,Collins:2002kn}. More recently, a similar sign change test has been proposed for the gluon Sivers function as well~\cite{Boer:2016fqd,Boer:2015ika}. Experimental verification of these properties would strongly corroborate our present understanding of the structure of the proton  and nonperturbative QCD effects.

Among gluon TMDs, the distribution of linearly polarized gluons inside an unpolarized proton~\cite{Mulders:2000sh,Meissner:2007rx,Boer:2016xqr} has attracted a lot of attention in the last few years. It corresponds to an interference between $+1$ and $-1$ gluon helicity states which, if sizable,  can affect the transverse momentum distributions of final state particles like, for instance, the Higgs boson~\cite{Boer:2011kf,Boer:2013fca,Echevarria:2015uaa}. Linearly polarized gluons have been investigated theoretically in the dilute-dense regime in proton-nucleus and lepton-nucleus collisions as well~\cite{Dominguez:2010xd,Metz:2011wb,Dominguez:2011br,Akcakaya:2012si,Schafer:2013wca,Dumitru:2015gaa,Marquet:2017xwy,Dominguez:2011wm}. 
Very interestingly, it turns out that at small-$x$ fractions of the gluons inside a nucleus, the linearly polarized distribution may reach its maximally allowed size, bounded by the unpolarized gluon density~\cite{Mulders:2000sh}, although it depends on the process whether the observable effects are maximal \cite{Boer:2017xpy}.

From the experimental point of view, almost nothing is known about gluon TMDs, because they typically require higher-energy scattering processes and are harder to isolate as compared to quark TMDs. Many proposals have been put forward to access them by looking at transverse momentum distributions and azimuthal asymmetries for bound or open heavy-quark pair production, both in lepton-proton and in proton-proton collisions. The reason is that heavy quarks are very sensitive to the gluon content of hadrons, as  is well known from studies of gluon PDFs. A first Gaussian shape extraction of the unpolarized TMD gluon distribution has been recently performed from LHCb data on the transverse spectra of $J/\psi$ pairs~\cite{Lansberg:2017dzg}. 

In a series of papers~\cite{Boer:2010zf,Pisano:2013cya,Boer:2016fqd}, the process $e\, p \to e^\prime \,Q \,\overline{Q}\, X$, with $Q$ being either a charm or a bottom quark, has been considered as a tool to extract gluon TMDs at a future Electron-Ion Collider (EIC)~\cite{Boer:2011fh,Accardi:2012qut,Zheng:2018awe}. The observables, needed to disentangle the five different gluon TMDs contributing to the unpolarized and transversely polarized cross sections, have been properly defined, each one of them corresponding to a specific azimuthal modulation. Moreover, especially in Ref.~\cite{Boer:2016fqd}, attention has been paid to the small-$x$ behavior of all the distributions and to their process dependence, by relating them to other reactions which could be measured, for example, at the proposed fixed target experiment AFTER@LHC~\cite{Brodsky:2012vg,Hadjidakis:2018ifr}. It is natural at this point to perform a similar analysis for the case in which the two heavy quarks form a bound state. We therefore consider here inclusive $J/\psi$ and $\Upsilon$ production in deep-inelastic lepton-proton scattering, namely  $e\,p \to e^\prime \,J/\psi \,(\Upsilon)\,X$, where the electron is unpolarized and the proton can be either unpolarized,  or polarized transversely to the electron-proton plane. In addition to unpolarized quarkonium production, we examine the cases in which the quarkonium state is polarized either longitudinally or transversely with respect to its direction of motion in the $\gamma^*p$ center-of-mass frame, with $\gamma^*$ being the virtual photon exchanged in the reaction.
Analogous studies, although limited to the Sivers and linearly polarized gluon densities and to unpolarized quarkonium production, have been published recently~\cite{Mukherjee:2016qxa,Rajesh:2018qks}. 

In the present analysis we adopt the TMD framework in combination with nonrelativistic QCD (NRQCD)~\cite{Hagler:2000dd,Yuan:2000qe,Yuan:2008vn}, which is the effective field theory that allows for a factorized treatment of the heavy-quark pair production, calculable in perturbative QCD, and the nonperturbative hadronization process leading to the binding of the pair, encoded in long-distance matrix elements (LDMEs)~\cite{Bodwin:1994jh}. Since these LDMEs, which are assumed to be universal, obey specific scaling rules in the average velocity $v$ of the heavy quark in the quarkonium rest frame~\cite{Lepage:1992tx}, the corresponding cross section can be evaluated through a double expansion in the strong coupling constant $\alpha_s$  and in the velocity $v$, with $v^2 \simeq 0.3$ for charmonium and $v^2 \simeq 0.1$ for bottomonium. In general, a heavy quark-antiquark pair can be produced in a color-singlet (CS) configuration, with the same quantum numbers as the observed quarkonium, but also as a color-octet state (CO) with different quantum numbers. In the latter case, the pair becomes colorless after the emission of soft gluons. The CS LDMEs are commonly obtained from potential models~\cite{Eichten:1995ch}, lattice calculations~\cite{Bodwin:1996tg} or from leptonic decays~\cite{Bodwin:2007fz}, while the CO ones are usually determined by fits to data on $J/\psi$ and $\Upsilon$ yields~\cite{Butenschoen:2010rq,Chao:2012iv,Sharma:2012dy,Bodwin:2014gia,Zhang:2014ybe}, but not from lattice calculations. As a result, at present our knowledge of the CO matrix elements is not very accurate (cf.\ Tables \ref{tab:jpsiLDME} and \ref{tab:uLDME} below). Moreover, although NRQCD successfully explains many experimental observations, it has problems to reproduce  all cross sections and polarization measurements for charmonia in a consistent way~\cite{Brambilla:2010cs,Andronic:2015wma}. As a consequence, alternative approaches to NRQCD are used as well, also in TMD studies. For instance, $J/\psi$ photoproduction as a way to access the gluon Sivers function~\cite{Godbole:2012bx,Godbole:2013bca,Godbole:2014tha} has been studied in the so-called Color Evaporation Model~\cite{Fritzsch:1977ay}, which is based on quark-hadron duality and assumes that the probability to form a physical (colorless) quarkonium state does not depend on the color and the other quantum numbers of the hadronizing $Q \overline{Q} $ pair. 
f
The TMD framework is based on TMD factorization, which, while not proven specifically for the process $e\,p \to e^\prime \,{\cal Q}\,X$ with ${\cal Q}=J/\psi\, (\Upsilon)$, has been rigorously proven for the analogous SIDIS process $e\,p \to e^\prime \, h \,X$, with $h$ a light hadron~\cite{Collins:2011zzd}. At leading order, they differ by the underlying hard process, which is  $\gamma^* q$ scattering in the latter versus $\gamma^* g$ scattering in the former. However, this does not make a difference from the perspective of TMD factorization, and neither does the mass of the final state hadron. Therefore, we expect TMD factorization to hold in $e\,p \to e^\prime \,{\cal Q}\,X$, in the kinematical configuration $P_{\mathcal{Q}T} \ll M_{\cal Q}$ and $Q \sim M_{\cal Q}$. Moreover, in this process and in these kinematics, the CO production mechanism is expected to be the dominant one~\cite{Fleming:1997fq,Sun:2017wxk}. In addition, some of our proposed observables, namely the single spin asymmetries, are expected to vanish in semi-inclusive deep-inelastic scattering in the CS mechanism, due to the absence of any initial or final state interactions~\cite{Yuan:2008vn}. Most of the previous studies on gluon TMDs in proton-proton collisions focussed on scattering processes in which the CS production mechanism is the dominant one, such as $ p\,p\to \eta_{c,b}\, X$, $p\,p\to \chi_{0c, b} \,(\chi_{2 c,b})\,X $~\cite{Boer:2012bt,Echevarria:2019ynx}, $p\,p\to J/\psi\,(\Upsilon)\,\gamma\, X$~\cite{Dunnen:2014eta},  $p\,p\to J/\psi\, (\Upsilon)\,\ell\, \bar \ell \,X$~\cite{Lansberg:2017tlc} and  $p\,p\to J/\psi\, J/\psi\,X$~\cite{Lansberg:2017dzg,Scarpa:2019fol}. The reason to concentrate on these CS dominated processes is to avoid the presence of final state interactions which, together with the initial state interactions present in proton-proton collisions, would lead to the breaking of TMD factorization~\cite{Rogers:2010dm}. Furthermore, as already discussed in 
Ref.~\cite{Boer:2016fqd}, the gluon distributions extracted in $ e\,p \to e^\prime \,J/\psi \,(\Upsilon)\,X $  or in $e\, p \to e^\prime \,Q \,\overline{Q}\, X $, which correspond to the so-called Weizs\"acker-Williams (WW) distributions in the small-$x$ limit, are all related to the TMDs entering in the above mentioned proton-proton reactions and differ from them by, at most, an overall minus sign. 

Investigating the process  $ e\,p \to e^\prime \,J/\psi \,(\Upsilon)\,X$ can be very helpful to improve our understanding of the mechanisms underlying quarkonium production. To this end, here we propose a new method to extract, apart from various TMDs that are at present still unknown from the experimental point of view, also the dominant CO LDMEs, namely $\langle0\vert{\cal O}_{8}^{J/\psi}(^{1}S_{0})\vert0\rangle$ and $\langle0\vert{\cal O}_{8}^{J/\psi}(^{3}P_{0})\vert0 \rangle$, by combining measurements of azimuthal asymmetries in $ e\,p \to e^\prime \,J/\psi \,(\Upsilon)\,X$, with analogous ones in $e\, p \to e^\prime \,Q \,\overline{Q}\, X$. In this way, heavy-quark final states at an EIC can contribute to the determination of the CO LDMEs. Here a complicating factor is the transition from the CO $Q \overline{Q} $ state into the true CS hadronic final state by means of soft gluon radiation (which resembles fragmentation into a light hadron) about which nothing quantitative is known, as far as we know. As a first step we consider this transition as infinitely narrow, i.e.\ as a delta function in transverse momentum (like often done for jets), but we also study the effect of smearing numerically\footnote{This can be viewed as a model study of the additional TMD shape function of Ref.~\cite{Echevarria:2019ynx}, which is considered as the TMD extension of the LDMEs.}. In addition, in order to avoid having to deal with evolution in the comparison of the two processes, one should consider the same value of the photon virtuality $Q^2$ in both processes. In the first one, $ e\,p \to e^\prime \,J/\psi \,(\Upsilon)\,X$, we consider the transverse momentum $P_{\mathcal{Q}T}$ of the produced quarkonium small with respect to the quarkonium mass $M_{\cal Q}\approx 2 M_Q$. In order to avoid the presence of two very different hard scales, we take $Q= 2 M_{ Q}$. Although one can take the same $Q$ value in the second process, $e\, p \to e^\prime \,Q \,\overline{Q}\, X$, there will be another hard scale given by the transverse momentum $K_\perp$ of each heavy quark, which we assume to be $K_\perp=Q=2 M_Q$ for simplicity.

The processes considered in this paper are gluon induced, and are therefore expected to be enhanced when considering smaller $x$ values. At an EIC, the smaller the $x$ value, the smaller the $Q$ values covered, so one has to keep a balance between the $x$ and $Q$ ranges. For the $J/\psi$ case one can go to lower $x$ values. Since we consider only a limited $Q$ range, we will not include TMD evolution, although this can be done along the lines considered in Ref.~\cite{Mukherjee:2016qxa}. To assess the less studied influence of evolution in $x$, we perform a numerical study of the implications nonlinear small-$x$ evolution would have in the range from $x\sim10^{-2}$ to $ x\sim10^{-4}$ covered by the EIC at low $Q$ values of a few GeV. It turns out to have only a moderate suppression effect. This study is limited to the unpolarized proton case, for which nonperturbative models are available for the corresponding small-$x$ gluon distributions~\cite{McLerran:1993ni,McLerran:1993ka,McLerran:1994vd}, which we use as the initial condition for the evolution. The Color Glass Condensate effective theory~\cite{Gelis:2010nm} makes it then possible to calculate the nonlinear evolution in rapidity of these distributions, in the presence of saturation. This was done with the help of a numerical implementation of JIMWLK on the lattice in Refs.~\cite{Dumitru:2015gaa,Marquet:2016cgx,Marquet:2017xwy}.  We use the results therein obtained for the unpolarized and linearly polarized WW TMDs inside an unpolarized hadron, to show predictions for our azimuthal modulations at different values of rapidity in the low-$x$ limit.

The paper is organized as follows. In Section~\ref{sec:def} we provide the operator definition of gluon TMDs and discuss their process dependence. The derivation of the cross section for unpolarized quarkonium production in DIS, within the TMD framework, can be found in Section~\ref{sec:HQ}. Further details of the calculations are relegated to Appendix~\ref{sec:app}. The azimuthal moments providing direct access to gluon TMDs are defined in Section~\ref{sec:asymm}. Similar observables for polarized quarkonium production are discussed in Section~\ref{sec:pol}.  Our strategy for the extraction of the CO LDMEs, based on the combination of azimuthal asymmetries for bound and open heavy-quark pair production, is described in Section~\ref{sec:LDME}, followed by a numerical study of smearing effects on this extraction in Section~\ref{sec:smearing}. Upper limits of the azimuthal moments, as well as an analysis of the small-$x$ evolution of gluon TMDs and the $\cos 2\phi$ asymmetries, are presented in Section~\ref{sec:pheno}. Summary and conclusions are given in Section~\ref{sec:conc}.

\section{Operator definition of gluon TMDs}
\label{sec:def}

The transverse momentum distribution of a gluon with four-momentum $p$ inside a proton with four-momentum $P$ and spin vector $S$ can be defined as follows. We first perform a Sudakov decomposition of $p$ and $S$ in terms of $P$ and a light-like vector $n$, conjugate to $P$.
Namely, 
\begin{align}
p^\mu & = x\,P^\mu  + p_\sT^\mu  + p^- n^\mu\, ,\\
S^\mu & = \frac{S_L}{M_p}\, \bigg (  P^\mu -  \frac{M_p^2}{P\cdot n}\, n^\mu\bigg )  + S_\sT^\mu\,,
\end{align}
where $M_p$ is mass of the proton and $S_\sT^2 = -\bm S_\sT^2$, with $0 \le S_L^2, \bm S_\sT^2 \le 1$,  such that $S_L^2 + \bm S_\sT^2= 1$. We then introduce the following matrix element of a correlator of the gluon field strengths $F^{\mu \nu}(0)$ and $F^{\nu \sigma}(\xi)$, evaluated at fixed light-front (LF) time $\xi^+ =\xi{\cdot}n=0$,
\begin{eqnarray}
\label{GluonCorr}
 {\Gamma}_g^{\mu\nu}(x,\bm p_\sT )
& = &  \frac{n_\rho\,n_\sigma}{(P{\cdot}n)^2}
{\int}\frac{\mathrm{d}(\xi{\cdot}P)\,\mathrm{d}^2\xi_\sT}{(2\pi)^3}\ e^{ip\cdot\xi}\, \langle P,
S|\,\tr\big[\,F^{\mu\rho}(0)\, U_{[0,\xi]} F^{\nu\sigma}(\xi)\,U^\prime_{[\xi, {0}]}\,\big] \,|P, S
\rangle\,\big|_{\text{LF}}\,,
\label{eq:g-corr}
\end{eqnarray}
with  $U_{[0,\xi]}$ and $U^\prime_{[0,\xi]}$ being two process dependent gauge links (or Wilson lines) that are needed to ensure gauge invariance. By means of the symmetric and antisymmetric transverse projectors, respectively  given by 
\begin{align}
g^{\mu\nu}_{\sT} & = g^{\mu\nu} - P^{\mu}n^{\nu}/P{\cdot}n-n^{\mu}P^{\nu}/P{\cdot}n \,,\\
\epsilon_\sT^{\mu\nu}  & = \epsilon^{\alpha\beta\mu\nu} P_\alpha n_\beta/P\cdot n\,, \quad  \text{with} \quad \epsilon_\sT^{1 2} = +1 \,, 
\end{align}
the correlator in Eq.~\eqref{GluonCorr} can be parametrized in terms of gluon TMDs~\cite{Mulders:2000sh,Meissner:2007rx,Boer:2016xqr}. 
For an unpolarized proton,  one has
\begin{align}
 {\Gamma}_U^{\mu\nu}(x,\bm p_\sT )  = & \frac{x}{2}\,\bigg \{-g_\sT^{\mu\nu}\,f_1^g (x,\bm p_\sT^2) +\bigg(\frac{p_\sT^\mu p_\sT^\nu}{M_p^2}\,
    {+}\,g_\sT^{\mu\nu}\frac{\bm p_\sT^2}{2M_p^2}\bigg) \,h_1^{\perp\,g} (x,\bm p_\sT^2) \bigg \} \,,
\label{eq:PhiparU}    
\end{align}
where $f_1^g (x,\bm p_\sT^2)$ is the TMD unpolarized distribution and  $h_1^{\perp\,g} (x,\bm p_\sT^2) $ is the distribution of linearly polarized gluons. They are both $T$-even, i.e.\ they can be  nonzero even in those processes where there are neither initial nor final state interactions. The correlator for a transversely polarized proton can be parametrized in terms of five independent gluon TMDs as follows
\begin{align} 
 {\Gamma}_T^{\mu\nu}(x,\bm p_\sT )   = & \frac{x}{2}\,\bigg \{g^{\mu\nu}_\sT\,
    \frac{ \epsilon^{\rho\sigma}_\sT p_{\sT \rho}\, S_{\sT\sigma}}{M_p}\, f_{1T}^{\perp\,g}(x, \bm p_\sT^2) + i \epsilon_\sT^{\mu\nu}\,
    \frac{p_\sT \cdot S_\sT}{M_h}\, g_{1T}^{g}(x, \bm p_\sT^2) \nonumber \\
    &  \,  + \,  \frac{p_{\sT \rho}\,\epsilon_\sT^{\rho \{ \mu}p_\sT^{\nu \}}}{2M_p^2}\,\frac{p_\sT\cdot S_\sT }{M_p} \, h_{1 T}^{\perp\,g}(x, \bm p_\sT^2)\,- \,\frac{p_{\sT \rho} \epsilon_\sT^{\rho \{ \mu}S_\sT^{\nu \}}\, + \,
      S_{\sT\rho} \epsilon_\sT^{\rho \{ \mu } p_\sT^{\nu \}}}{4M_p} \, h_{1T}^{g}(x, \bm p_\sT^2) \,\,\bigg \}\,,
\label{eq:Phipar}
\end{align}
where the symmetrization operator is defined as $p^{ \{ \mu} q^{\nu \} }=p^\mu q^\nu + p^\nu q^\mu$.
The three gluon TMDs that appear in its symmetric part, $(\Gamma_T^{\mu\nu}+\Gamma_T^{\nu\mu})/2$,  are all $T$-odd:  
$f_{1T}^{\perp\,g}(x, \bm p_\sT^2)$ is the gluon Sivers function, while   the $h$ functions are chiral-even distributions of linearly polarized gluons inside a transversely polarized proton. 
In analogy to the transversity function for quarks,  we define the combination
\begin{equation}
h_1^g \equiv h_{1T}^g +\frac{\bm p_\sT^2}{2 M_p^2}\,  h_{1T}^{\perp\,g}\,,
\label{eq:h1}
\end{equation}
which however, in contrast to quark transversity, vanishes upon integration over transverse momentum~\cite{Boer:2016fqd}.  

Because of the definition in Eq.~\eqref{eq:g-corr}, the TMDs introduced in Eqs.~\eqref{eq:PhiparU} and \eqref{eq:Phipar}  will depend on the gauge links, the specific structure of which is determined by the process under consideration. In this case, as in $e\, p \to e^\prime \, Q\, \overline{Q}\, X$~\cite{Boer:2016fqd}, the partonic reaction $\gamma^* g \to Q \,\overline Q$ probes gluon TMDs with two future pointing Wilson lines, denoted as  $+$ links. In the small-$x$ limit they correspond to the WW distributions. As already pointed out in 
Ref.~\cite{Boer:2016fqd}, these TMDs can be related to the ones having two past-pointing, or $-$ gauge links, which could be accessed in processes like  $p\,  p\to \gamma \, \gamma \, X$ in the back-to-back correlation limit~\cite{Qiu:2011ai}. More specifically, the $T$-even unpolarized and linearly polarized gluon TMDs are expected to be the same in the two kind of processes, while the $T$-odd densities, like the gluon Sivers functions, should be related by a minus sign. On the other hand,  gluon TMDs with both a $+$ and $-$ link (future and past
pointing),  corresponding to the dipole distributions at small $x$, cannot be related to the TMDs discussed here.  They could be accessed in processes like $p\,  p\to \gamma^* \, \text{jet}\, X$ \cite{Dominguez:2011wm}, in the kinematic region where gluons in the polarized proton dominate, such that the partonic channel $q \, g \to \gamma^* \, q$ is effectively selected~\cite{Boer:2017xpy}. However, TMD factorization for $p\,  p\to \gamma^* \, \text{jet}\, X$ has not been established so far.

\section{Outline of the calculation}
\label{sec:HQ}
\begin{figure}[t]
\begin{centering}
\includegraphics[trim={0cm 23cm 0cm 3cm},clip,scale=1]{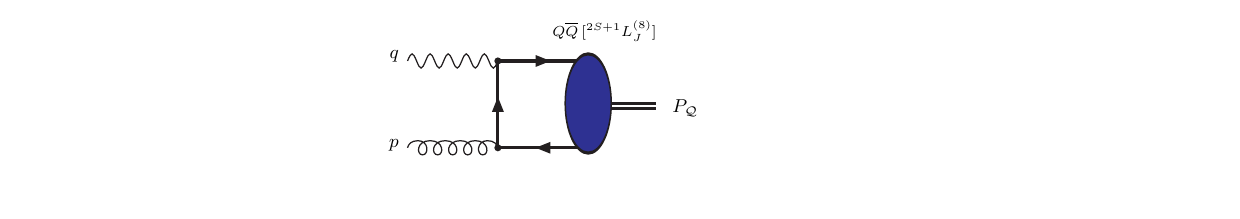}
\par\end{centering}
\caption{Leading order diagram for  the process $\gamma^* (q) \,+ \, g(p)\to {\cal Q} (P_{\cal Q})$, with ${\cal Q} = J/\psi$ or $\Upsilon$. The crossed diagram, in which the directions of the arrows are reversed, is not shown. Only the color-octet configurations $^1S_0^{(8)}$, $^3P^{(8)}_{J}$ with $J=0,1,2$, contribute, as it turns out from the calculation described in Appendix~\ref{sec:app}.}
\label{fig:fd-lo}
\end{figure}

We study the process
\begin{equation}
e(\ell) + p(P,S) \to e(\ell^{\prime}) + {\cal Q}\,(P_{\cal Q}) + X\,,
\end{equation}
where ${\cal Q}$ is either a $J/\psi$ or a $\Upsilon$ meson, the incoming proton is polarized with polarization vector $S$, and the other particles are unpolarized. We choose the  reference frame such that both the virtual photon exchanged in the reaction and the incoming proton move along the $\hat z$-axis, and azimuthal angles are measured w.r.t.\ to the lepton scattering plane, such that $\phi_\ell = \phi_\ell^\prime =0$. Moreover, in order to apply a framework based on TMD factorization, we consider only the kinematic region in which the component of the quarkonium momentum  transverse w.r.t.\ the lepton plane, denoted by $q_\sT \equiv  P_{{\cal Q} \sT}$, is small compared to the virtuality of the photon $Q$ and to the mass of the quarkonium $M_{\cal Q}$. The differential cross section can be written as 
\begin{eqnarray}
\d\sigma
& = &\frac{1}{2 s}\,\frac{\d^3 \ell'}{(2\pi)^3\,2 E_e^{\prime}} \frac{\d^3 P_{\cal Q}}{(2\pi)^3\,2 E_{\cal Q}}
{\int}\d x\, \d^2\bm p_{\sT}\,(2\pi)^4
\delta^4(q {+} p {-} P_{\cal Q})
 \nonumber \\
&&\qquad \qquad\qquad \qquad\qquad\times  \frac{1}{x^2\,Q^4}\, 
L_{\mu \rho}(\ell,q) \,  \Gamma_{g\,\nu\sigma}(x{,} \bm p_{\sT})
\, H^{\mu\nu}_{\gamma^*\, g \rightarrow {\cal Q} } \,H^{\star\, \rho\sigma}_{\gamma^*\, g \rightarrow {\cal Q} } \, , 
\label{CrossSec}
\end{eqnarray}
where $s =  (\ell + P)^2 \approx 2\,\ell\cdot P$ is the total invariant mass squared and  $Q^2 = -q^2 \equiv -(\ell-\ell^\prime)^2$. Moreover, the gluon correlator $ \Gamma_g$ is defined in Eq. \eqref{GluonCorr} and the leptonic tensor $L(\ell, q)$ is given by
\begin{equation}
L^{\mu\nu}(\ell, q) = e^2 \left [-g^{\mu\nu}\,Q^2 + 2\,( 
 \ell^{\mu}\ell'^\nu + \ell^{\nu}\ell'^\mu) \right ]\, ,
\end{equation}
with $e$ the electric charge of the electron. 

The calculation proceeds along the same lines of Ref.~\cite{Pisano:2013cya}, which we summarize for completeness in the following.  We start with introducing the light-like vectors $n_+$ and $n_-$, which obey the relations  $n_+^2 = n_-^2 = 0$ and $n_+ \cdot n_- =1$.
Then we note that the four-momenta $P$ and $q$ can be written as
\begin{equation}
P = n_+ + \frac{M_p^2}{2}\,n_- \approx n_+
\quad \mbox{and} \quad
q = -\xB\,n_+ + \frac{Q^2}{2\,\xB}\,n_- \approx -\xB\,P + (P\cdot q)\,n_-\,,
\label{eq:qexpansion}
\end{equation}
where  $x_B$ is the Bjorken-$x$ variable, with $\xB = Q^2/2P\cdot q$ up to target mass corrections.
We will thus perform a Sudakov decomposition of all the momenta in the reaction in terms of $n_+ = P$ and $n_- = n = (q+\xB\,P)/P\cdot q$.
Therefore, the  leptonic momenta can be written as
\begin{eqnarray}
\ell & = &\frac{1-y}{y}\,\xB\,P + \frac{1}{y}\,\frac{Q^2}{2\xB}\,n
+ \frac{\sqrt{1-y}}{y}\,Q\,\hat\ell_\perp\, ,
\\
\ell^\prime & = & \frac{1}{y}\,\xB\,P + \frac{1-y}{y}\,\frac{Q^2}{2\xB}\,n
+ \frac{\sqrt{1-y}}{y}\,Q\,\hat\ell_\perp\,,
\end{eqnarray}
where we have introduced the inelasticity variable  $y = P\cdot q/P\cdot \ell$, such that the following relations hold: $s = 2\,P\cdot q/y = Q^2/\xB y$. 
The invariant mass squared of the virtual photon-target
system is defined as $W^2 =(q+P)^2$, and can be expressed  in terms of the other invariants: $W^2= Q^2(1-\xB)/\xB = (1-\xB)ys$. Similarly, the gluon momentum can be expanded as
\begin{equation}
p =x\,P + p_\sT + (p\cdot P-x\,M_p^2)\,n \approx x\,P + p_\sT\,,
\label{PartonDecompositions}
\end{equation}
where $x = p\cdot n$, while for the momentum of the quarkonium state ${\cal Q}$ we have
\begin{eqnarray}
P_{\cal Q}
&=& z \,(P\cdot q)\,n + \frac{M_{\cal Q}^2 + \bm P_{{\cal Q}\sT}^2}{2z \,P\cdot q}\,P + P_{{\cal Q}\sT}\,,
\label{eq:jetmom1}
\end{eqnarray}
with $z= P_{\cal Q}\cdot P/ q \cdot P$  and $P_{{\cal Q}\sT}^2 = -\bm P_{{\cal Q}\sT}^2$. 

In a reference frame in which azimuthal angles are measured w.r.t.\ the lepton plane ($\phi_{\ell}=\phi_{\ell^\prime}=0$),
denoting by $\phi_S$, $\phi_\sT$  the azimuthal angles of the three-vectors $\bm S_\sT$ and $\bm P_{\mathcal{Q}_\sT}$,  respectively, 
the phase-space elements in  Eq.~(\ref{CrossSec}) can be written as
\begin{equation}
\frac{\d^3 \ell'}{(2\pi)^3\, 2 E_e^{\prime}} 
= \frac{1}{16 \pi^2}\, {s}{y} \,\d\xB\,\d y\, ,
\quad \mbox{and} \quad 
\frac{\d^3 P_{\cal Q}}{(2\pi)^3\, 2 E_{\cal Q}} = \frac{1}{2 (2\pi)^3}\,  \frac{\d z}{z}\,\d^2 \bm P_{\mathcal{Q}_\sT}\,.
\end{equation}
Furthermore,  using the Sudakov decomposition of the gluon  momentum in Eq.\ \eqref{PartonDecompositions}, 
the $\delta$-function in Eq.~(\ref{CrossSec}) can be re-expressed  as
\begin{equation}
\label{DeltaFunc}
\delta^4(p+q-P_{\cal Q})
=\frac{2}{y \,s}\, \delta\bigg(x-\xB -\frac{M_{\cal Q}^2}{y \,z\,s}\bigg)
\,\delta (1-z) 
\,\delta^2\left (\bm p_\sT-\bm P_{\mathcal{Q}_\sT}\right )\,.
\end{equation}
Therefore, upon integration over the variables $x$, $z$ and $\bm p_\sT$, the cross section takes the final form
\begin{equation}
\frac{\d\sigma}
{\\d y\,\d\xB\,\d^2\bm{q}_{\sT}} \equiv \d\sigma (\phi_S, \phi_\sT) =    \d\sigma^U(\phi_\sT)  +  \d\sigma^T (\phi_S, \phi_\sT)  \,,
\label{eq:cs}
\end{equation}
with $z$ fixed to the value $z=1$, the transverse momentum of the incoming gluon equal to that of the quarkonium, and its
longitudinal momentum fraction $x$ given by
\begin{align}
x  &  = \xB + \frac{M^2_{\cal Q}}{y\,s}  = \frac{M^2_{\cal Q} + Q^2}{y\,s} =    \xB\,\frac{M^2_{\cal Q} + Q^2}{Q^2} \, .
\label{eq:xresult}
\end{align}

Within the framework of NRQCD, at leading order in the strong coupling constant $\alpha_s$, the partonic subprocess that contributes to $J/\psi$ production is  $\gamma^*g \to Q \overline Q  [ ^{2S+1}L_J^{(8)} ] $, as depicted in Fig.~\ref{fig:fd-lo}, where we have used a spectroscopic notation to indicate that the $Q \overline Q$ pair forms a bound state with  spin $S$, orbital angular momentum $L$ and total angular momentum $J$. The additional superscript $(8)$ denotes the color configuration. The relevant CO LDMEs are  $\langle 0 \vert {\cal O}_8^{J/\psi} (^1 S_0)\vert 0 \rangle$ and $\langle 0 \vert {\cal O}_8^{J/\psi} (^3 P_J)\vert 0 \rangle$, with $J=0,1,2$. The CS production mechanism is possible only at 
${\cal O}(\alpha_s^2)$, where the $Q \overline Q$ is formed at short distances in a $^3S_1^{(1)}$ configuration in association with a gluon. As pointed out in Ref.~\cite{Fleming:1997fq}, the CS contribution is suppressed relatively to the CO by a perturbative coefficient of the order $\alpha_s/\pi$. On the other hand,  $\langle 0 \vert {\cal O}_8^{J/\psi} (^1 S_0)\vert 0 \rangle$ and $\langle 0 \vert {\cal O}_8^{J/\psi} (^3 P_J)\vert 0 \rangle$ are suppressed as compared to $\langle 0 \vert {\cal O}_1^{J/\psi} (^3 S_1)\vert 0 \rangle$ by $v^3$ and $v^4$,  respectively.  Hence, according to the NRQCD scaling rules, the CO contribution should be enhanced by about a factor $v^3 \pi/\alpha_s \approx 2$ with respect to the CS one. This factor becomes $\approx 4$ in the actual numerical analysis presented in Ref.~\cite{Fleming:1997fq} for values of  $Q^2 > 4$ GeV$^2$. A further suppression of the CS contribution can be achieved by applying  a cut on the variable $z$, for instance by taking $z \ge 0.9$,  because at high $z$ the CS term is known to become negligible~\cite{Fleming:1997fq} and will be therefore neglected in our analysis. 
Of course, since the true final state quarkonium must really be a color singlet, the transition from the $Q\overline{Q}$ pair into the quarkonium state is an idealization in the sense that we take it as a delta function in transverse momentum space. 
Within these approximations, the final unpolarized and transversely polarized cross sections read
\begin{align}
\d\sigma^U
  & =  {{\cal N}}\, \bigg [ A^U  f_1^g (x, \bm q_\sT^2 )+  \frac{\bm q_\sT^2}{M_p^2}\, B^U\, h_1^{\perp\, g} (x, \bm q_\sT^2 ) \cos 2 \phi_\sT  \bigg ] \,,
\label{eq:csU}
\end{align}
and
\begin{align}
\d\sigma^T & =   {\cal N}\,\vert \bm S_\sT\vert \,\frac{\vert\bm q_\sT\vert }{M_p} \bigg \{A^T f_{1T}^{\perp\,g} (x, \bm q_\sT^2 ) \sin(\phi_S -\phi_\sT)  + B ^T \,\left [ h_{1}^{g} (x, \bm q_\sT^2 )\, \sin(\phi_S+\phi_\sT)  \,- \,\frac{\bm q_\sT^2}{2 M_p^2}\, h_{1\sT}^{\perp\,g} (x, \bm q_\sT^2 )\, \sin (\phi_S - 3 \phi_\sT)  \right ] \bigg \} \, ,
\label{eq:csT}
\end{align}
with the normalization factor  $\cal N$ given by
\begin{equation}
{\cal N} =   (2 \pi)^2  \frac{{\alpha^2 \alpha_se_Q^2}}{ y\,  Q^2\, M_{\cal Q}(M_{\cal Q}^2+Q^2)} \,,
\label{eq:N}
\end{equation}
where $e_Q$ is the fractional electric charge of the quark $Q$.
Details of the derivation can be found in Appendix~\ref{sec:app}. 
Expressions for $A^U, B^{U}$ and $A^T$ have also been given in Ref.~\cite{Mukherjee:2016qxa}, where some power suppressed terms were included, as well as an additional power suppressed $\cos\phi_T$ amplitude.
The explicit expressions of the terms $A^{U/T}$ in Eqs.~(\ref{eq:csU}) and (\ref{eq:csT}) read
\begin{align}
{A}^U  = A^T
  = &~  [1+(1-y)^2]\,{\cal A}_{U+L}^{\gamma^*g\to {\cal Q}}  \, - \,   y^2\, {\cal A}_{L}^{\gamma^*g \to {\cal Q}} \,,
\label{eq:Agstar}  \\
{B}^U =  {B}^T =
 &  ~  (1-y)\, {\cal B}_{T}^{\gamma^*g \to {\cal Q}}\,, 
\end{align} 
where the subscripts $U+L$, $L$, $T$ refer to the specific polarization of the photon~\cite{Pisano:2013cya,Brodkorb:1994de}. If we denote by ${\cal A}_{\lambda_\gamma,\lambda_\gamma^\prime}$,  with $\lambda_\gamma, \lambda_\gamma^\prime =0, \pm 1$, 
the  helicity amplitudes squared for the process $\gamma^* g \to Q \overline Q \big [^{2 S +1}L^{(8)}_J \big ]$, the following relations hold (omitting numerical prefactors)
\begin{eqnarray}
{\cal A}_{U+L} &  \propto & {\cal A}_{++} + {\cal A}_{--} + {\cal A}_{00}\, ,
\nonumber \\
{\cal A}_{L} &  \propto & {\cal A}_{00}\,, \nonumber \\
{\cal A}_{I} &  \propto & {\cal A}_{0+} +{\cal A}_{+0}-{\cal A}_{0-} -{\cal A}_{-0} \,, \nonumber \\
{\cal A}_{T} &  \propto & {\cal A}_{+-} + {\cal A}_{-+}~.
\label{eq:Ahel}
\end{eqnarray}
Furthermore, in terms of the CO LDMEs we obtain 
\begin{align}
{\cal A}_{U+L}^{\gamma^* g\to {\cal Q}} \,  = & \,  \langle 0 \vert {\cal O}_8^{J/\psi} (^1 S_0)\vert 0 \rangle +\frac{4}{N_c}\,\frac{1}{M_{\cal Q}^2 (M_{\cal Q}^2+Q^2)^2} \left  [( 3M_{\cal Q}^2 +Q^2)^2\, \langle 0 \vert {\cal O}_8^{J/\psi} (^3 P_0)\vert 0 \rangle
 \right . \nonumber  \label{eq:AUL}\\
& \qquad \qquad \qquad +\ 2 \,Q^2(2M_{\cal Q}^2+Q^2) \,\langle 0 \vert {\cal O}_8^{J/\psi} (^3 P_1)\vert 0 \rangle  + \left . \frac{2}{5}\,(6 M_{\cal Q}^4 + 6 M_{\cal Q}^2 \,Q^2 + Q^4)\,  \langle 0 \vert {\cal O}_8^{J/\psi} (^3 P_2)\vert 0 \rangle  \right ]\,,\\
    {\cal A}_{L}^{\gamma^*g \to {\cal Q}} = &  \frac{16}{N_c}\,
   \frac{Q^2}{(M_{\cal Q}^2+Q^2)^2} \left [ \,\langle 0 \vert {\cal O}_8^{J/\psi} (^3 P_1)\vert 0 \rangle  +  \frac{3}{5}\,
 \langle 0 \vert {\cal O}_8^{J/\psi} (^3 P_2)\vert 0 \rangle \right ] \,,
 \label{eq:AL}
\end{align}
 and
\begin{align}
{\cal B}_{T}^{\gamma^*g \to {\cal Q}}  = &- \langle 0 \vert {\cal O}_8^{J/\psi} (^1 S_0)\vert 0 \rangle +\frac{4}{N_c} \, \frac{1}{M_{\cal Q}^2(M_{\cal Q}^2+Q^2)^2} \left [ ( 3M_{\cal Q}^2 +Q^2)^2 \, \langle 0 \vert {\cal O}_8^{J/\psi} (^3 P_0)\vert 0 \rangle \right .\nonumber \\ 
&\qquad \qquad \qquad \left . - 2\, Q^4 \,\langle 0 \vert {\cal O}_8^{J/\psi} (^3 P_1)\vert 0 \rangle 
 +\frac{2}{5}\,{Q^4}\,  \langle 0 \vert {\cal O}_8^{J/\psi} (^3 P_2)\vert 0 \rangle \right ]\, .
 \label{eq:BT}
\end{align}
We conclude this section by noticing that each of the four independent azimuthal modulations in the cross section for $e\,p\,\to e^\prime \,J/\psi\,X$, that is  $\cos 2 \phi_\sT$, $\sin(\phi_S-\phi_\sT)$, $\sin(\phi_S+\phi_\sT)$ and $\sin(\phi_S-3\phi_\sT)$, probe a different gluon TMD.  These modulations are the same as for the process $e\,p\to e^\prime\,Q\,\overline Q\,X$~\cite{Boer:2016fqd}, after integration over the azimuthal angle $\phi_\perp$. As already pointed out in Ref.~\cite{Boer:2016fqd}, such angular structures and the corresponding TMDs are very similar to the  quark asymmetries in the SIDIS process $e\, p \to e^\prime \, h\, X$, where the role of $\phi_\sT$ is played by $\phi_h$~\cite{Boer:1997nt}. 

\section{Azimuthal asymmetries}
\label{sec:asymm}

In order to single out the different azimuthal modulations of the cross section $\d\sigma$, given in Eq.~\eqref{eq:cs} and Eqs.~\eqref{eq:csU}-\eqref{eq:csT}, we define the following azimuthal moments
\begin{align}
A^{W(\phi_S,\phi_\sT)} & \equiv 2\,\frac {\int   \d \phi_S \, \d \phi_\sT \, W(\phi_S,\phi_\sT)\,\d\sigma (\phi_S,\,\phi_\sT)}{\int  \d \phi_S\,  \d \phi_\sT \,\d\sigma (\phi_S, \phi_\sT)}\,, 
\label{eq:mom}
\end{align}
where the denominator reads 
\begin{align}
\int  \d \phi_S\, \d \phi_\sT \,\d\sigma (\phi_S,\phi_\sT)  & \equiv   \int  \d \phi_S\,\d \phi_\sT \,\frac{\d\sigma^U}{\d y\,\d\xB\,\d^2 \bm{q}_{\sT} } =   (2 \pi)^2\, {\cal N}\, A^U\,f_1^g(x, \bm q_\sT^2)
\end{align}
with $\cal N$ and $ A^U$ given by Eqs.~\eqref{eq:N} and \eqref{eq:Agstar}, respectively.  By taking $W= \cos  2\phi_\sT$ we obtain   
\begin{align}
\langle \cos 2\phi_\sT \rangle & \equiv \frac{1}{2}\, A^{\cos 2 \phi_\sT}  =  \frac{(1-y)\, {\cal B}_{T}^{\gamma^*g \to {\cal Q }} }{[1+(1-y)^2]{\cal A}^{\gamma^*g \to {\cal Q}} _{U+L} -y^2 {\cal A}_L^{\gamma^*g \to {\cal Q }} }\, \frac{ \bm q_\sT^2 }{2 M_p^2}\, \frac{h_{1}^{\perp\,g}(x,\bm q_\sT^2)}{f_1^g(x,\bm q_\sT^2)}\,.
\label{eq:A0}
\end{align}
Moreover, assuming $\vert \bm S_\sT \vert = 1$, the other moments can be written as
\begin{align}
A^{\sin(\phi_S-\phi_\sT)} & =  \frac{\vert \bm q_\sT\vert}{M_p}\, \frac{f_{1T}^{\perp\,g}(x,\bm q_\sT^2)}{f_1^g(x,\bm q_\sT^2)}\,, 
\label{eq:A1}\\
A^{\sin(\phi_S+\phi_\sT)}  & = 
\frac{(1-y)\, {\cal B}_{T}^{\gamma^*g \to {\cal Q}} }{[1+(1-y)^2]{\cal A}^{\gamma^*g \to {\cal Q }} _{U+L} -y^2 {\cal A}_L^{\gamma^*g \to {\cal Q } }}
 \,\frac{\vert \bm q_\sT\vert}{M_p}\, \frac{h_{1}^{g}(x,\bm q_\sT^2)}{f_1^g(x,\bm q_\sT^2)}\,,
 \label{eq:A2}\\
A^{\sin(\phi_S-3\phi_\sT)}  & =   - \frac{(1-y)\, {\cal B}_{T}^{\gamma^*g \to {\cal Q }} }{[1+(1-y)^2]{\cal A}^{\gamma^*g \to {\cal Q}} _{U+L} -y^2 {\cal A}_L^{\gamma^*g \to {\cal Q }} }\, \frac{\vert \bm q_\sT\vert^3}{2 M_p^3}\, \frac{h_{1\sT}^{\perp \,g}(x,\bm q_\sT^2)}{f_1^g(x,\bm q_\sT^2)}\label{eq:A3}\,,
\end{align}
where the explicit expressions for  ${\cal A}_{U+L}^{\gamma^{*}g\to{\cal Q}}$, ${\cal A}_{L}^{\gamma^{*}g\to{\cal Q}}$,  ${\cal B}_{T}^{\gamma^{*}g\to{\cal Q}}$ in Eqs.~\eqref{eq:AUL}-\eqref{eq:BT} can be further simplified, if one employs the heavy-quark spin symmetry relations~\cite{Bodwin:1994jh}  
\begin{equation}
 \langle 0 \vert {\cal O}_8^{J/\psi} (^3 P_J)\vert 0 \rangle=(2J+1)\langle 0 \vert {\cal O}_8^{J/\psi} (^3 P_0)\vert 0 \rangle\, + \,{{\cal O}(v^2)}\,. 
 \label{eq:hqss}
 \end{equation}
 Hence, at leading order in $v$, we obtain:
\begin{align}
{\cal A}_{U+L}^{\gamma^{*}g\to{\cal Q}}= & \langle0\vert{\cal O}_{8}^{J/\psi}(^{1}S_{0})\vert0\rangle+\frac{12}{N_{c}}\frac{7M_{{\cal Q}}^{2}+3Q^{2}}{M_{{\cal Q}}^{2}(M_{{\cal Q}}^{2}+Q^{2})}\langle0\vert{\cal O}_{8}^{J/\psi}(^{3}P_{0})\vert0\rangle\,,\\
{\cal A}_{L}^{\gamma^{*}g\to{\cal Q}}= & \frac{96}{N_{c}}\,\frac{Q^{2}}{(M_{{\cal Q}}^{2}+Q^{2})^{2}}\langle0\vert{\cal O}_{8}^{J/\psi}(^{3}P_{0})\vert0\rangle\,,
\end{align}
and
\begin{align}
{\cal B}_{T}^{\gamma^{*}g\to{\cal Q}}= & -\langle0\vert{\cal O}_{8}^{J/\psi}(^{1}S_{0})\vert0\rangle+\frac{12}{N_{c}}\frac{3 M_{{\cal Q}}^{2}-Q^{2}}{M_{{\cal Q}}^{2}(M_{{\cal Q}}^{2}+Q^{2})}\, \langle0\vert{\cal O}_{8}^{J/\psi}(^{3}P_{0})\vert0\rangle\, .
\end{align}

The asymmetries in Eqs.~(\ref{eq:A0}), (\ref{eq:A2}) and (\ref{eq:A3}) vanish in the limit $y\to 1$ when the virtual photon is longitudinally polarized. Moreover, very importantly, we point out that a measurement of the ratios 
\begin{align}
\frac{A^{\cos 2\phi_\sT }}{A^{\sin(\phi_S+\phi_\sT)}}= \frac{\bm q_\sT^2}{M_p^2}\, \frac{h_{1}^{\perp\,g}(x,\bm q_\sT^2)}{h_{1}^{g}(x,\bm q_\sT^2)}\,,
\label{eq:ratioA1}\\
\frac{A^{\sin(\phi_S-3\phi_\sT)}}{A^{\cos 2\phi_\sT }}=- \frac{\vert \bm q_\sT\vert}{2 M_p}\, \frac{h_{1 \sT}^{\perp\,g}(x,\bm q_\sT^2)}{h_{1}^{\perp\,g}(x,\bm q_\sT^2)}\,,\\
\frac{A^{\sin(\phi_S-3\phi_\sT)}}{A^{\sin(\phi_S+\phi_\sT)}} =- \frac{\bm q_\sT^2}{2 M_p^2}\, \frac{h_{1T}^{\perp\,g}(x,\bm q_\sT^2)}{h_{1}^{g}(x,\bm q_\sT^2)}\, 
\label{eq:ratioA3}
\end{align}
would directly probe the relative magnitude of the different gluon TMDs, without any dependence on the color octet LDMEs. Notice that 
Eqs.~\eqref{eq:ratioA1}-\eqref{eq:ratioA3} are not based on the heavy-quark spin symmetry relations in Eq.~\eqref{eq:hqss}. 

It would be interesting to check experimentally the behavior of these ratios of asymmetries because currently there are no reliable theoretical predictions. However, for the {\it dipole} gluon TMDs we expect, from model independent considerations~\cite{Boer:2016xqr}, that  the observable in Eq.~\eqref{eq:ratioA3} will reach the value one in the small-$x$ limit. It remains to be seen if this holds also for the WW gluon distributions discussed in this paper. 

\section{Quarkonium polarization}
\label{sec:pol}

The study of $J/\psi$  polarization is often considered as a test of NRQCD. Hence, in this section we present  the cross sections for the processes $e \, p\, \to \, e^\prime\, {\cal Q}_{L/T}\, X$, where the quarkonium in the final state  is polarized either longitudinally ($L$) or transversely ($T$) with respect to the direction of its three-momentum in the photon-proton center-of-mass frame. The cross sections have the same angular structure as in Eq.~\eqref{eq:cs} and Eqs.~\eqref{eq:csU}-\eqref{eq:csT}. Namely, in terms of the kinematic variables defined in the previous section, 
\begin{equation}
\frac{\d\sigma^P}
{\\d y\,\d\xB\,\d^2\bm{q}_{\sT}} \equiv \d\sigma^P (\phi_S, \phi_\sT) =    \d\sigma^{UP}(\phi_\sT)  +  \d\sigma^{TP} (\phi_S, \phi_\sT)  \,,
\label{eq:csP}
\end{equation}
where the superscript $P=L $ or $T$ denotes the polarization of the quarkonium and  the superscripts $U$ and $T$ refer to the possible polarization states of the initial proton. Clearly, $\d\sigma = \d\sigma^L+\d\sigma^T$, where $\d\sigma$ is the cross section for unpolarized quarkonium  production given in Eq.~\eqref{eq:cs}. Furthermore, 
\begin{align}
\d\sigma^{U P}
  & =  {{\cal N}}\, \bigg [ A^{UP}  f_1^g (x, \bm q_\sT^2 )+  \frac{\bm q_\sT^2}{M_p^2}\, B^{UP}\, h_1^{\perp\, g} (x, \bm q_\sT^2 ) \cos 2 \phi_\sT  \bigg ] \,,
\label{eq:csUP}
\end{align}
and
\begin{align}
\d\sigma^{TP} & =   {\cal N}\,\vert \bm S_\sT\vert \,\frac{\vert\bm q_\sT\vert }{M_p} \bigg \{A^{TP} f_{1T}^{\perp\,g} (x, \bm q_\sT^2 ) \sin(\phi_S -\phi_\sT)  + B ^{TP} \,\left [ h_{1}^{g}\, \sin(\phi_S+\phi_\sT)  \,- \,\frac{\bm q_\sT^2}{2 M_p^2}\, h_{1\sT}^{\perp\,g}\, \sin (\phi_S - 3 \phi_\sT)  \right ] \bigg \} \, ,
\label{eq:csTP}
\end{align}
with $\cal N$ defined in Eq.~\eqref{eq:N} and 
\begin{align}
{A}^{UP}  = A^{TP}
  = &~  [1+(1-y)^2]\,{\cal A}_{U+L}^{\gamma^*g\to {\cal Q}_P}  \, - \,   y^2\, {\cal A}_{L}^{\gamma^*g \to {\cal Q}_P} \,,
\label{eq:AgstarP}  \\
{B}^{UP} =  {B}^{TP} =
 &  ~  (1-y)\, {\cal B}_{T}^{\gamma^*g \to {\cal Q}_P}\,.
\end{align} 
The explicit expressions for longitudinally polarized  quarkonium production read
\begin{align}
{\cal A}_{U+L}^{\gamma^{*}g\to{\cal Q}_L}= &\, \frac{1}{3}\, \langle0\vert{\cal O}_{8}^{J/\psi}(^{1}S_{0})\vert0\rangle+\frac{12}{N_{c}}\frac{M_{{\cal Q}}^{4}+ 10 \, M_{\cal Q}^2\,  Q^2+Q^{4}}{M_{{\cal Q}}^{2}(M_{{\cal Q}}^{2}+Q^{2})^2}\langle0\vert{\cal O}_{8}^{J/\psi}(^{3}P_{0})\vert0\rangle\,,
\label{eq:AUL_L}\\
{\cal A}_{L}^{\gamma^{*}g\to{\cal Q}_L}=  &\,   {\cal A}_{L}^{\gamma^{*}g\to{\cal Q}} = \frac{96}{N_{c}}\,\frac{Q^{2}}{(M_{{\cal Q}}^{2}+Q^{2})^{2}}\langle0\vert{\cal O}_{8}^{J/\psi}(^{3}P_{0})\vert0\rangle\,, 
\end{align} 
in agreement with the results in Ref.~\cite{Fleming:1997fq}, while 
\begin{align}
{\cal B}_{T}^{\gamma^{*}g\to{\cal Q}_L}= & \, -\frac{1}{3}\, \langle0\vert{\cal O}_{8}^{J/\psi}(^{1}S_{0})\vert0\rangle+\frac{12}{N_{c}}\frac{1}{M_{{\cal Q}}^{2}} \, \langle0\vert{\cal O}_{8}^{J/\psi}(^{3}P_{0})\vert0\rangle\, 
\end{align}
is new. For completeness, the results corresponding to transverse polarization of the quarkonium read
\begin{align}
{\cal A}_{U+L}^{\gamma^{*}g\to{\cal Q}_T}= &\, \frac{2}{3}\, \langle0\vert{\cal O}_{8}^{J/\psi}(^{1}S_{0})\vert0\rangle+\frac{24}{N_{c}}\frac{3\, M_{{\cal Q}}^{4}+Q^{4}}{M_{{\cal Q}}^{2}(M_{{\cal Q}}^{2}+Q^{2})^2}\langle0\vert{\cal O}_{8}^{J/\psi}(^{3}P_{0})\vert0\rangle\,,\\
{\cal A}_{L}^{\gamma^{*}g\to{\cal Q}_T}=  &\,  0\, ,
\end{align}
and
\begin{align}
{\cal B}_{T}^{\gamma^{*}g\to{\cal Q}_T}= & \, -\frac{2}{3}\, \langle0\vert{\cal O}_{8}^{J/\psi}(^{1}S_{0})\vert0\rangle+ \frac{24}{N_{c}}\frac{1}{M_{{\cal Q}}^{2}} \,  \frac{M_{\cal Q}^2-Q^2}{M_{\cal Q}^2+Q^2} \, \langle0\vert{\cal O}_{8}^{J/\psi}(^{3}P_{0})\vert0\rangle\, .
\end{align}
More details on the derivation of the above cross sections, performed along the lines of Refs.~\cite{Beneke:1996yw,Braaten:1996jt,Leibovich:1996pa}, can be found at the end of  Appendix \ref{sec:app}.

We note that, also for polarized quarkonium production, it is possible to define azimuthal moments exactly as in Eq.~\eqref{eq:mom}, as well as  their ratios in Eqs.~\eqref{eq:ratioA1}-\eqref{eq:ratioA3}.  In particular, it turns out that such ratios of asymmetries  depend neither on the  LDMEs,  nor on the polarization state of the detected quarkonium. 

\section{A strategy for the determination of the dominant CO LDME\MakeLowercase{s}}
\label{sec:LDME}
In this section we define novel observables, which within our approximations are only sensitive to the CO LDMEs $\langle0\vert{\cal O}_{8}^{J/\psi}(^{1}S_{0})\vert0\rangle$ and $\langle0\vert{\cal O}_{8}^{J/\psi}(^{3}P_{0})\vert0\rangle$, and to the corresponding ones for $\Upsilon$ production,  but not to TMDs. This is possible by combining azimuthal asymmetries in $ e\,p \to e^\prime \,J/\psi \,(\Upsilon)\,X$ with analogous quantities for open heavy-quark pair production in $e\, p \to e^\prime \,Q \,\overline{Q}\, X$~\cite{Boer:2010zf,Pisano:2013cya,Boer:2016fqd} in the following way, 
\begin{align}
{\cal R}^{\cos2\phi_\sT} & = \frac{\int  \d \phi_\sT \cos 2\phi_\sT\, \d\sigma^{{\cal Q}} (\phi_S,\phi_\sT)}{\int  \d \phi_\sT \, \d\phi_\perp \cos 2\phi_\sT\, \d\sigma^{Q\overline Q} (\phi_S,\phi_\sT, \phi_\perp) }\,,
\label{eq:R2phi}\\
{\cal R} & = \frac{\int  \d \phi_\sT \, \d\sigma^{{\cal Q}} (\phi_S,\phi_\sT)}{\int  \d \phi_\sT \, \d\phi_\perp\, \d\sigma^{Q\overline Q} (\phi_S,\phi_\sT, \phi_\perp) }\, ,
\label{eq:R}
\end{align}
where $\d\sigma^{\cal Q}$ now denotes the differential cross section for the process $ e\,p \to e^\prime\,{\cal Q}\,X$ defined in Eq.~\eqref{eq:cs}, and 
\begin{equation}
 \d\sigma^{Q\overline Q} \equiv \frac{\d\sigma^{Q \overline Q}}{\d z \,\d y\,\d\xB\,\d^2 \bm{K}_{\perp} \,\d^2 \bm{q}_{\sT} }
\end{equation}
is the differential cross section for the process 
\begin{equation}
 e(\ell) \,+\,p (P,S)\,\to \,e^\prime(\ell^\prime) \,+\, Q(K_Q)\,+\,\overline{Q}(K_{\overline Q})\,+\,X,
 \end{equation}
 in which the quark-antiquark pair is almost back-to-back in the plane orthogonal to the direction of the proton and the exchanged virtual photon. Hence, in the $\gamma^*p$ center-of-mass frame, the difference of the transverse momenta of the outgoing quark and  antiquark, conventionally specified by $\bm K_\perp =(\bm K_{Q\perp}-\bm K_{\overline Q \perp})/2$, should be large compared to their sum $\bm q_\sT = \bm K_{Q\perp}+\bm K_{\overline Q \perp}$. In Eqs.~\eqref{eq:R2phi}-\eqref{eq:R}, $\phi_S$, $\phi_\sT$ and $\phi_\perp$  are the azimuthal angles of the proton polarization vector $\bm S$, $\bm q_\sT$ and $\bm K_\perp$, respectively. Furthermore,  $y$ is the inelasticity and $x_B$ is the Bjorken variable,  while $z = K_Q \cdot  P/ q\cdot P$, with $q = \ell-\ell^\prime$ as for  $ e\,p \to e^\prime\,{\cal Q}\,X$. We note that in the definitions of ${\cal R}^{\cos2\phi_\sT}$ and ${\cal R}$ the proton does not need to be polarized.

The hard scale of the process  $ e\,p \to e^\prime \,{\cal Q}\,X$ is identified with the quarkonium mass $M_{\cal Q}\approx 2 M_Q$. Moreover, to avoid the presence of two very different hard scales in the calculation of the numerators of ${\cal R}^{\cos2\phi_\sT}$ and ${\cal R}$, we simply take the photon virtuality $Q$ to be $Q=M_{\cal Q} \approx 2 M_Q$. Therefore, from the results for the cross section presented above, we obtain
\begin{align}
\int  \d \phi_\sT \cos 2\phi_\sT\, \d\sigma^{{\cal Q}} (\phi_S,\phi_\sT)  & \equiv   \int \d \phi_\sT \cos 2\phi_\sT\,\frac{\d\sigma^U}{\d y\,\d\xB\,\d^2 \bm{q}_{\sT} } =   \pi\, {\cal N}\, B^U\, \frac{\bm q_\sT^2}{ M_p^2}\,h_1^{\perp\, g}(x, \bm q_\sT^2) \nonumber \\
& = \pi ^3 \, \frac{\alpha^2 \alpha_s e^2_Q}{ 16\, M_Q^5} \left ( \frac{1-y}{y} \right  ) \, \left [-{\cal O}_8^S + \frac{1}{M_Q^2} \, {\cal O}_8^P \right ]\,  \frac{\bm q_\sT^2}{ M_p^2}\,h_1^{\perp\, g}(x, \bm q_\sT^2)\,, \label{eq:c2phi-Jpsi}\\
\int  \d \phi_\sT \,\d\sigma^{{{\cal Q}}} (\phi_S,\phi_\sT)  & \equiv   \int \d \phi_\sT \,\frac{\d\sigma^U}{\d y\,\d\xB\,\d^2 \bm{q}_{\sT} } =   2 \pi\, {\cal N}\, A^U\,f_1^g(x, \bm q_\sT^2)\nonumber \\
& = \pi^3 \,   \frac{\alpha^2 \alpha_s e^2_Q}{ 8\, M_Q^5}  \,\left [ \frac{1+(1-y)^2}{y} \, {\cal O}_{8}^S+ \frac{10-10y+3y^2 }{y}\,
\frac{1}{M_{Q}^{2}}{\cal O}_{8}^P  \right ] \,f_1^g(x, \bm q_\sT^2)\,,
\label{eq:A-Jpsi}
\end{align}
where we have introduced the shorthand notation $ {\cal O}_{8}^S \equiv  \langle0\vert{\cal O}_{8}^{{\cal Q}}(^{1}S_{0})\vert0\rangle $  and $ {\cal O}_{8}^P \equiv  \langle0\vert{\cal O}_{8}^{{\cal Q}}(^{3}P_{0})\vert0\rangle $.

\begin{table}
\begin{centering}
\begin{tabular}{|c|c|c|c|}
\hline 
$J/\psi$ & $\langle0|\mathcal{O}_{8}^{J/\psi}\bigl(^{1}S_{0}\bigr)|0\rangle$ & $\langle0|\mathcal{O}_{8}^{J/\psi}\bigl(^{3}P_{0}\bigr)|0\rangle/M_c^2$ & \tabularnewline
\hline 
\hline 
CMSWZ~\cite{Chao:2012iv}
& $8.9 \pm 0.98$ & $0.56 \pm 0.21$ & $\times10^{-2}\,\mathrm{GeV}^{3}\mathrm{}$\tabularnewline
SV~\cite{Sharma:2012dy}& $1.8 \pm 0.87 $ & $1.8 \pm 0.87$ & $\times10^{-2}\,\mathrm{GeV}^{3}$\tabularnewline
BK~\cite{Butenschoen:2010rq}
& $4.50 \pm 0.72$ & $-1.21 \pm 0.35$ & $\times10^{-2}\,\mathrm{GeV}^{3}\mathrm{}$\tabularnewline
BCKL~\cite{Bodwin:2014gia}
& $9.9 \pm 2.2$ & $  1.1\pm 1.0$ & $\times10^{-2}\,\mathrm{GeV}^{3}\mathrm{}$\tabularnewline
\hline
\end{tabular}
\par\end{centering}
\caption{Numerical values of the LDMEs for $J/\psi$ production.}
\label{tab:jpsiLDME}
\vspace{1cm}

\begin{centering}
\begin{tabular}{|c|c|c|c|}
\hline 
$\Upsilon\bigl(1S\bigr)$ & $\langle0|\mathcal{O}_{8}^{\Upsilon(1S)} (^{1}S_{0}  )|0\rangle$ & $\langle0|\mathcal{O}_{8}^{\Upsilon (1S )}\bigl(^{3}P_{0}\bigr)|0\rangle/(5M_b^2)$ & \tabularnewline
\hline 
\hline 
SV~\cite{Sharma:2012dy}& $1.21 \pm 4.0$ & $1.21 \pm 4.0$ & $\times10^{-2}\,\mathrm{GeV}^{3}$\tabularnewline
\hline 
\end{tabular}
\par\end{centering}
\caption{Numerical values of the LDMEs for $\Upsilon(1S)$ production.}
\label{tab:uLDME}
\end{table}

Since we would like to have an exact cancellation of the gluons TMDs in the ratio, we need to consider the same value of the photon virtuality $Q$ in both processes.  Furthermore, the other hard scale $K_\perp \equiv \vert \bm K_\perp \vert$ in $e\, p \to e^\prime \,Q \,\overline{Q}\, X$ is taken to be  $K_\perp=Q$ to avoid any possible TMD evolution effect. From the results in Ref.~\cite{Boer:2016fqd} calculated at  $K_\perp=Q = 2 M_Q$ and $z = 1/2$, we get
\begin{align}
\int  \d \phi_\sT \, \d \phi_\perp \cos 2\phi_\sT\, \d\sigma^{Q \overline Q} (\phi_S,\phi_\sT\, \phi_\perp)  & \equiv   \int \d \phi_\sT\, \d \phi_\perp\cos 2\phi_\sT\,\frac{\d\sigma^{Q \overline Q}}{\d z\, \d y\,\d\xB\,\d^2 \bm{K}_{\perp}\d^2 \bm{q}_{\sT} }
\label{eq:c2phi-QQb}
 \nonumber \\
& =   - \pi \, \frac{\alpha^2 \alpha_s e^2_Q}{108 \, M_Q^4}  \left ( \frac{1-y }{y} \right )  \frac{\bm q_\sT^2}{ M_p^2}\,h_1^{\perp\, g}(x, \bm q_\sT^2)\,,  \\
\int  \d \phi_\sT \, \d\phi_\perp \, \d\sigma^{Q \overline Q} (\phi_S,\phi_\sT, \phi_\perp)  & \equiv   \int \d \phi_\sT \, \d \phi_\perp\,\frac{\d\sigma^{Q \overline Q}}{\d z \,\d y\,\d\xB\,\d^2 \bm{K}_{\perp} \,\d^2 \bm{q}_{\sT} } 
\nonumber \\
& =  \pi \, \frac{\alpha^2 \alpha_s e^2_Q}{54 \, M_Q^4} \,\left (  \frac{26 -26 y + 9y^2}{y} \right ) \,f_1^g(x, \bm q_\sT^2)\,.
\label{eq:A-QQb}
\end{align}
By taking the ratio of the two $\cos2\phi_\sT$-weighted cross sections in Eqs.~\eqref{eq:c2phi-Jpsi} and \eqref{eq:c2phi-QQb}, and the  ratio of 
the two cross sections in Eqs.~\eqref{eq:A-Jpsi} and \eqref{eq:A-QQb}, it turns out that the two independent observables
\begin{align}
{\cal R}^{\cos2\phi_\sT} & = \frac{27 \pi^2}{4}\, \frac{1}{M_Q} \,  \left [{\cal O}_8^S - \frac{1}{M_Q^2} \, {\cal O}_8^P \right ] \,,\\
{\cal R} & = \frac{27\, \pi^2}{4}\,\frac{1}{M_Q} \, \frac{ [ 1+(1-y)^2 ] \, {\cal O}_{8}^S + ( 10-10y+3y^2  )\, {\cal O}_{8}^P /M_{Q}^{2}}{26 -26 y +9 y^2}  \,,
\label{eq:ratio-R}
\end{align}
give access to two different combinations of the CO LDMEs ${\cal O}_8^S $ and ${\cal O}_8^P$. Therefore the knowledge of both of them will allow to single out these two LDMEs, at least at leading order. TMDs will re-enter in higher orders in an {\it a priori} calculable way, thus introducing a $q_T$ dependence of ${\cal R}^{\cos2\phi_\sT}$ and ${\cal R} $ which at leading order and without final state smearing effects are constant.

A similar comparison could be done to jet pair production instead, but in that case quark contributions may spoil the cancellation of the gluon TMDs to some extent, depending on the kinematic region under study.
This may be an option worth considering, should open heavy-quark pair production turn out not to be feasible at an EIC. In a recent Monte Carlo analysis for a future EIC it is shown that at least single spin asymmetries will be hard to study in open heavy-quark pair production (assuming 10 fb$^{-1}$) \cite{Zheng:2018awe}, but here we propose a comparison to open heavy-quark pair production in the unpolarized proton case.

We point out that the measurement of quarkonium polarization will probe other combinations of the CO LDMEs, and can not only be used for consistency checks, but also to assess the importance of higher order contributions, which will be different for the unpolarized and polarized cases. By extending the definitions in Eqs.~\eqref{eq:R2phi} and \eqref{eq:R} to longitudinally polarized quarkonium production, using the cross sections presented Section~\ref{sec:pol}, we find 
\begin{align}
{\cal R}_L^{\cos2\phi_\sT} & =  \frac{27 \pi^2}{4}\, \frac{1}{M_Q} \,  \left [\frac{1}{3}\,{\cal O}_8^S - \frac{1}{M_Q^2} \, {\cal O}_8^P \right ] \,,\\
{\cal R}_L & = \frac{9\, \pi^2}{4}\,\frac{1}{M_Q} \, \frac{ [ 1+(1-y)^2 ] \, {\cal O}_{8}^S + 3\, ( 6-6y+y^2  )\, {\cal O}_{8}^P /M_{Q}^{2}}{26 -26 y +9 y^2}  \,,
\end{align}
while, for transversely polarized quarkonium production, 
\begin{align}
{\cal R}_T^{\cos2\phi_\sT}  & =   \frac{9\, \pi^2}{2}\, \frac{1}{M_Q} \,{\cal O}_8^S   \,,\\
{\cal R}_T & = \frac{9\, \pi^2}{2}\,\frac{1}{M_Q} \, \frac{ [ 1+(1-y)^2 ] \, {\cal O}_{8}^S + 3\, ( 2-2y+y^2  )\, {\cal O}_{8}^P /M_{Q}^{2}}{26 -26 y +9 y^2}  \,,
\end{align}
with
\begin{align}
{\cal R}_L^{\cos2\phi_\sT} + {\cal R}_T^{\cos2\phi_\sT} &  = {\cal R}^{\cos2\phi_\sT} \,,\\
{\cal R}_L + {\cal R}_T &  = {\cal R} \,.
\end{align}
In particular, we notice that the measurement of ${\cal R}_T^{\cos 2\phi_\sT}$,  for values of the photon virtuality such that $Q=M_{\cal Q}$, will directly probe the matrix element ${\cal O}_{8}^S=  \langle0\vert{\cal O}_{8}^{{\cal Q}}(^{1}S_{0})\vert0\rangle $.

Before moving on to the numerical studies, we like to comment on the robustness of the above results. As emphasized several times, the presented expressions are leading order (LO) in both TMD factorization and in NRQCD. Next-to-leading order (NLO) corrections will reintroduce sensitivity to the TMDs, but as mentioned the LO CO contribution is expected to be dominant over the NLO CS (and CO) contributions, parametrically by a factor $v^3 \pi/\alpha_s \approx 2$ and in practice by a larger factor for high $Q^2$~\cite{Fleming:1997fq}. As also mentioned, a further strong suppression of the NLO CS contribution can be achieved by applying a high lower-cut on the variable $z$, e.g.\ $z \ge 0.9$, as shown in \cite{Fleming:1997fq}. Furthermore, in a recent study \cite{Sun:2017wxk} it was shown that the NLO CS contribution undershoots the DIS data particularly at low transverse momenta. Both high $Q^2$ and low transverse momenta are considered in the present case in order to ensure TMD factorization of the process $e\,p \to e^\prime \,{\cal Q}\,X$. Despite the low transverse momenta ($P_{\mathcal{Q}T} \ll M_{\cal Q} \sim Q$), experimentally the quarkonium state should be clearly distinguishable from the proton remnants. This is unlike the case of proton-proton collisions, where the transverse momentum functions as a large scale.

Moreover, we would like to emphasize that our process is very much similar to $e \, p\, \to \, e^\prime \, \pi\,X$,  where $\gamma^* q \to q^\prime$ is the dominant channel, $Q$ is again the hard scale and the transverse momentum of the pion can be arbitrarily small. For such a process, in this kinematic configuration, a rigorous proof of TMD factorization exists~\cite{Collins:2011zzd}. The final state interactions of the fragmenting quark with the proton remnants can be summed up to yield a gauge link in the quark TMD correlator. The only difference with the  reaction under study,  $\gamma^* g \to {\cal Q}$,  is that the incoming parton is now a gluon and the final state interactions will be resummed in the gauge link of the gluon correlator, which will be in the adjoint representation instead of the fundamental one. The mass of the bound state Q does not affect the gauge link structure and hence no TMD factorization breaking problems due to color entanglement are present.

In short, the QCD corrections for this process in the kinematic region considered will not lead to a breaking of TMD factorization and are not expected to upset the NRQCD expansion and the CO dominance. Next we will study the effects of final state smearing due to the transition from the CO $Q \overline{Q} $ state into the true CS hadronic final state.  

\section{Smearing effects}
\label{sec:smearing}

In order to assess the impact of final state smearing we focus on the ratio ${\cal R}$ defined above, and introduce the functions $\Delta_L(\bm k_\sT^2)$, where $\bm k_\sT$ is the transverse momentum of the produced heavy quark-antiquark pair w.r.t.\ the CS hadronic final state. We assume that the smearing is different for the two color-octet states, identified by $L=0$ and $L=1$. Then Eq.~\eqref{eq:ratio-R} needs to be modified as follows
\begin{align}
{\cal R} & = \frac{27\, \pi^2}{4}\,\frac{1}{M_Q} \, \frac{ [ 1+(1-y)^2 ] \, {\cal O}_{8}^S \, S_0(x, \bm q_\sT^2)+ ( 10-10y+3y^2  )\, {\cal O}_{8}^P /M_{Q}^{2}\,S_1(x, \bm q_\sT^2) }{26 -26 y +9 y^2}  \,,
\end{align}
where
\begin{align}
S_L(x, \bm q_\sT^2) = \frac{{\cal C}[f_1^g \, \Delta_L] (x, \bm q_\sT^2)  }{f_1^g(x, \bm q_\sT^2)}\,,  \qquad{\text{with}} \qquad L=0,1\,,
\end{align}
and where we have introduced  the convolutions of  the TMD gluon distribution $f_1^g$ with $\Delta_L$, which are defined by
\begin{align}
{\cal C}[ f_1^g\,\Delta_L] (x, \bm q_\sT^2) \equiv  \int \d^2 \bm p_\sT \int \d^2 \bm k_\sT  \, \delta^2(\bm q_\sT - \bm p_\sT -\bm k_\sT)\, f_1^g(x, \bm p_\sT^2)\, \Delta_L(\bm k_\sT^2) \,.
\end{align}
The result in Eq.~\eqref{eq:ratio-R}  is recovered when 
\begin{align}
\Delta_0(\bm k_\sT^2) =\Delta_1(\bm k_\sT^2) = \delta^2(\bm k_\sT)\, ,
\end{align}
and, therefore, $S_0 = S_1=1$. 

We will adopt a model parameterization for the  transverse momentum dependent gluon distribution $f_1^g$, as it has not yet been extracted from experiments. For simplicity, the $x$ and $\bm p_\sT$ dependences are factorized:
\begin{align}
f_1^g(x, \bm p_\sT^2) =   \frac{C_\sT^2}{ 2\pi} \,f_1^g(x)\,  \frac{1}{1 + \bm p_\sT^2 \, C_\sT^2}   \, , 
\label{eq:input-TMD}
\end{align}
where $f_1^g(x)$ is the gluon distribution function integrated over $\bm q_\sT$, and $C_\sT$ is a constant whose value depends on the hard scale of the process. We assume $C_\sT^2= 4$ GeV$^{-2}$ for $J/\psi$ production and $C_\sT^2 = 1$  GeV$^{-2}$ for $\Upsilon$ production~\cite{Boer:2012bt}. This choice is motivated by TMD evolution, which is expected to make parton distributions flatter as the scale increases. This $q$-Gaussian or Tsallis distribution (also sometimes less accurately referred to as Gaussian+tail model) is considered more realistic than a pure Gaussian distribution, which falls off too fast.  Consequently, for Gaussians, $S_L$ will diverge for large $\bm q_\sT^2$ and require the inclusion of TMD evolution leading to a power-law fall off, as considered here. Note that for the adopted model the smearing functions $S_L$ do not depend on $x$, but only on $\bm q_\sT^2$. In principle, there may also be smearing in the denominator of $S_L$ due to the fragmentation of a heavy quark into a $D$- or $B$-meson, but since these are produced at large transverse momentum, this effect is expected to be much less relevant. 

To the best of our knowledge, no parametrization is so far available for the smearing functions $\Delta_L$. Therefore we propose a model based on the properties of the radial wave function of the hydrogen atom in momentum space, namely:
\begin{itemize}
\item For large $\bm p_\sT$, $\Delta_L$ vary as $(\bm p_\sT^2)^{-(L+4)}$, with $L=0,1$, independently of the heavy quark mass. 
\item For small $\bm p_\sT$,  $\Delta_L$ vary as $(\bm p_\sT^2)^L$, hence $\Delta_1$ vanishes at $\bm p_\sT =0$, while 
$\Delta_0$ does not. 
\end{itemize}
Furthermore, the normalization is fixed by imposing 
\begin{align}
\int \d^2 \bm k_\sT\,  \Delta_L (\bm k_\sT^2) = 1\,.
\end{align}
Explicitly we have
\begin{align}
\Delta_0 (\bm k_\sT^2) =   \frac{3 C_\sT^2}{ \pi} \,\frac{1}{(1 + \bm k_\sT^2 \, C_\sT^2)^4}    \,,\qquad \Delta_1 (\bm k_\sT^2) = \frac{12 C_\sT^4}{ \pi} \,\frac{\bm k_\sT^2}{(1 + \bm k_\sT^2 \, C_\sT^2)^5}   \,,
\end{align}
where $C_\sT$ is taken to be independent of $L$ and equal to the width of the TMD distribution in Eq.~\eqref{eq:input-TMD}. This guarantees that the transverse momentum distribution for a heavier quarkonium state falls off less fast, reflecting its smaller spatial extent.  

\begin{figure}[t]
\begin{centering}
\includegraphics[clip,scale=.7]{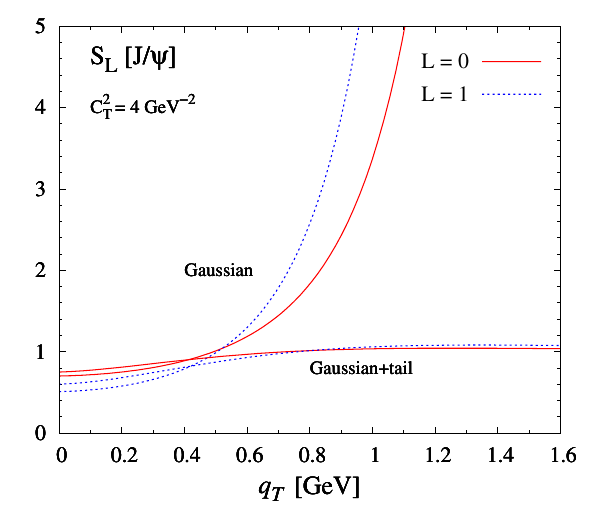}\includegraphics[clip,scale=.7]{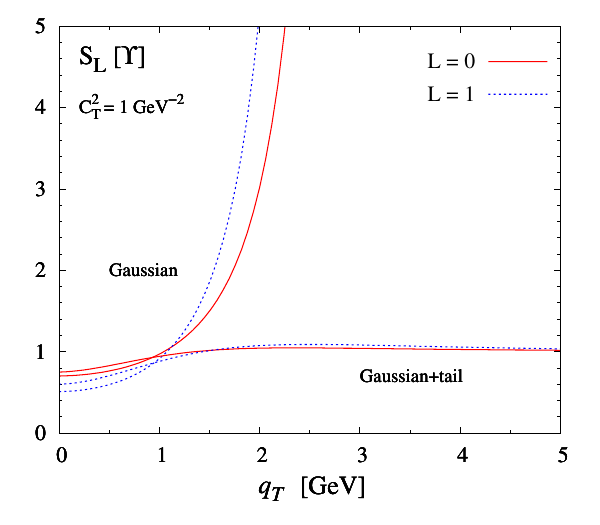}
\par\end{centering}
\caption{Transverse momentum dependence of the smearing functions $S_L$, with $L=0,1$,  for  $e \, p \to e^\prime \, J/\psi \, X $ (left panel) and $e \, p \to e^\prime \, \Upsilon\, X $ (right panel), for the adopted model parameterizations for the gluon TMD and $\Delta_L$ with the choices $C_\sT^2 = 4$ GeV$^{-2}$ ($J/\psi$ production) and  $C_\sT^2 = 1$ GeV$^{-2}$ ($\Upsilon$ production).}
\label{fig:smearing} 
\end{figure}

The transverse momentum dependence of the smearing functions  $S_L(\bm q_\sT^2)$, whose deviation from the value one is a signal of the presence of smearing effects, is shown in Fig.~\ref{fig:smearing} for both $J/\psi$ and $\Upsilon$ production. We note that the smearing effects are generally not sizable, except when $q_\sT$ is very close to zero.

We conclude that the observation of a $q_\sT$-dependence in the ratios ${\cal R}$ (and ${\cal R}^{\cos2\phi_\sT}$) indicates final state smearing and/or higher order effects. If indeed found to be moderate, this dependence can be included in the error on the extracted CO LDME values. 
The cross-check with the polarized quarkonium case can help further, because the smearing is expected to be the same in that case, as opposed to the higher order effects which moreover are calculable. In this way there should be sufficient experimental handles to test the validity of the approximations and estimate the uncertainties involved.

Regarding the effect of final state smearing on the presented azimuthal asymmetries, there is the problem that the gluon TMDs involved are entirely unknown. In the next section we therefore present upper bounds on the spin asymmetries for which we have explicitly checked that the smearing effects are small and comparable to those on ${\cal R}$, hence not considered important. Therefore, we will proceed with the delta function approximation in what follows.

\begin{figure}[htb]
\begin{centering}
\includegraphics[clip,scale=.7]{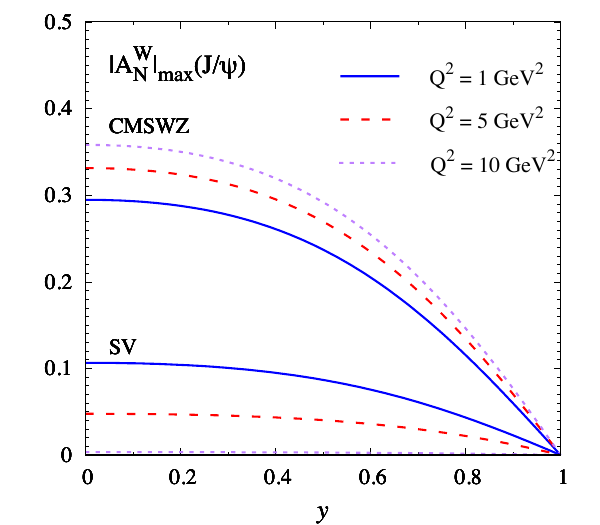}\includegraphics[clip,scale=.7]{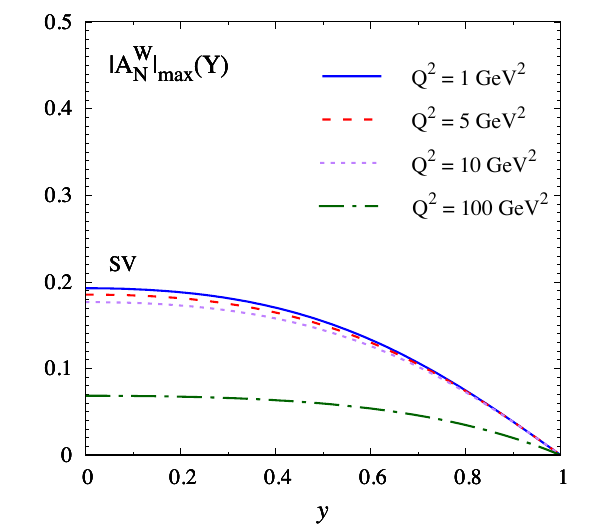}
\par\end{centering}
\caption{Upper bounds for the $\langle \cos 2\phi_\sT\rangle$ and  $A_N^W$  asymmetries, with $W =  \sin (\phi_S +\phi_\sT),\, \sin (\phi_S - 3 \phi_\sT)$, for $e \, p \to e^\prime \, J/\psi \, X $ (left panel) and  $e \, p \to e^\prime \, \Upsilon\, X $ (right panel), as a function of $y$ and for different values of the scale $Q^2$. The labels SV and CMSWZ denote the adopted LDME sets, given in Tables \ref{tab:jpsiLDME} and \ref{tab:uLDME}. The bounds obtained with the BCKL set, not shown explicitly, lie always between the SV and CMSWZ results presented in the left panel.}
\label{fig:cos2phisat} 
\end{figure}

{\section{Numerical results}
\label{sec:pheno}}

\subsection{Upper bounds of the asymmetries}
 
The polarized gluon TMDs have to satisfy the following, model independent, positivity bounds~\cite{Mulders:2000sh}
\begin{align}
\frac{\vert \bm p_\sT \vert }{M_p}\, \vert f_{1T}^{\perp \,g}(x,\bm p_\sT^2) \vert & \le   f_1^g(x,\bm p_\sT^2)\,,\nonumber \\
\frac{ \bm p^2_\sT }{2 M_p^2}\, \vert h_{1}^{\perp\,g}(x,\bm p_\sT^2) \vert & \le   f_1^g(x,\bm p_\sT^2)\,,\nonumber \\
\frac{\vert \bm p_\sT \vert }{M_p}\, \vert h_{1}^g(x,\bm p_\sT^2) \vert & \le   f_1^g(x,\bm p_\sT^2)\,,\nonumber \\
\frac{\vert \bm p_\sT \vert^3}{2 M_p^3}\, \vert h_{1T}^{\perp \,g}(x,\bm p_\sT^2) \vert & \le   f_1^g(x,\bm p_\sT^2)\,,
\label{eq:bounds}
\end{align}
which can be used to calculate the upper limits of the azimuthal moments defined in the previous section.  It can be easily seen 
that the Sivers asymmetry in Eq.~(\ref{eq:A1}) is bound to 1, while the asymmetries in Eqs.~(\ref{eq:A2}) and (\ref{eq:A3}) have the same upper bound, which we denote by $A_N^W$. This is also the same upper bound of the weighted cross section $\langle \cos2\phi_\sT\rangle$ defined in Eq.~\eqref{eq:A0}. 

For the numerical estimate of our asymmetries, we use different sets of extractions of the CO LDMEs for the $J/\psi$~\cite{Chao:2012iv,Sharma:2012dy,Butenschoen:2010rq,Bodwin:2014gia} (see Table~\ref{tab:jpsiLDME}), and one set for the $\Upsilon(1S)$ (Table~\ref{tab:uLDME}), all of them obtained from fits to TEVATRON, RHIC and LHC data.  Note that most of these results are obtained from NLO analyses, except for the SV set in Ref.~\cite{Sharma:2012dy}, which is based on a LO calculation like our asymmetries. The negative value of  $\langle0\vert{\cal O}_{8}^{J/\psi}(^{3}P_{0})\vert0\rangle$ from the BK set leads to negative LO unpolarized cross sections for certain values of $Q^2$. For this reason, the results obtained using the BK parametrization are not shown explicitly. The mass of the $J/\psi$ is taken to be 3.1 GeV, while the one for $\Upsilon$ is 9.5 GeV. We define the charm and bottom quark masses to be equal to half of the mass of the $J/\psi$ and the $\Upsilon$, respectively. 

\begin{figure}[t]
\begin{centering}
\includegraphics[clip,scale=.7]{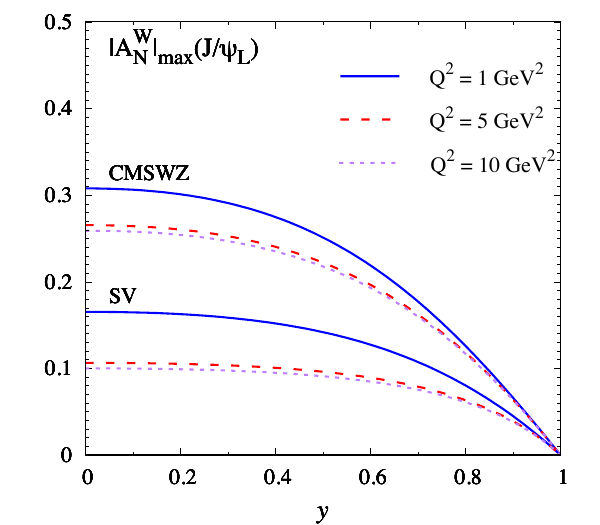}\includegraphics[clip,scale=.7]{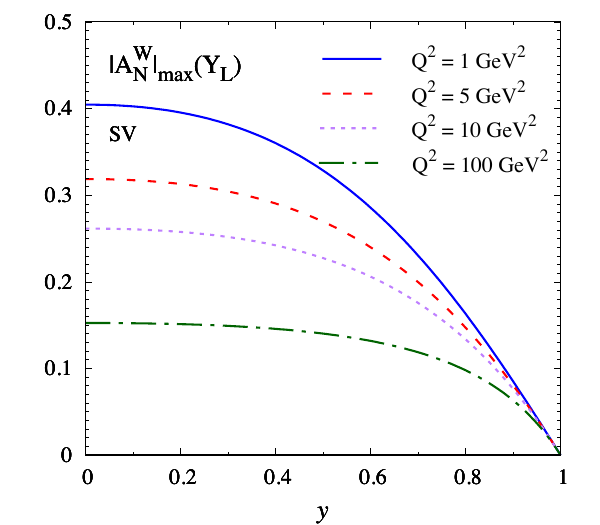}
\par\end{centering}
\caption{Upper bounds for the $\langle \cos 2\phi_\sT\rangle$ and  $A_N^W$  asymmetries, with $W =  \sin (\phi_S +\phi_\sT),\, \sin (\phi_S - 3 \phi_\sT)$, for  $e \, p \to e^\prime \, J/\psi \, X $ (left panel) and  $e \, p  \to e^\prime \, \Upsilon\, X $ (right panel), with the $J/\psi$ and $\Upsilon$ mesons longitudinally polarized along their direction of motion in the $\gamma^* p$ center-of-mass frame, as a function of $y$ and for different values of the scale $Q^2$. The labels SV and CMSWZ denote the adopted LDME sets, given in Tables \ref{tab:jpsiLDME} and \ref{tab:uLDME}. The bounds obtained with the BCKL set, not shown explicitly, lie always between the SV and CMSWZ results presented in the left panel.}
\label{fig:cos2phiLsat} 
\end{figure}

\begin{figure}[t]
\begin{centering}
\includegraphics[clip,scale=.7]{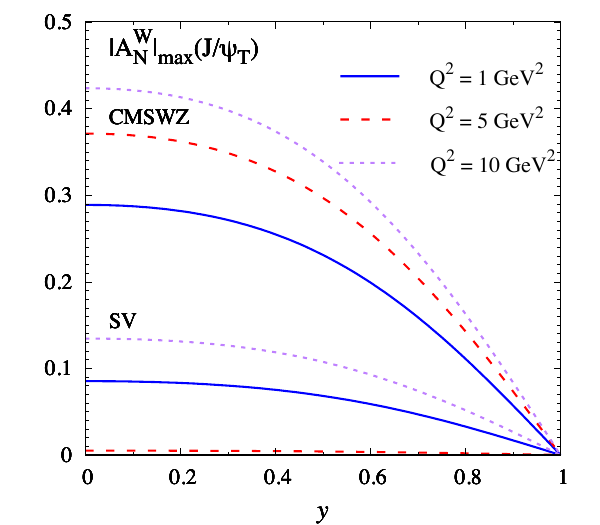}\includegraphics[clip,scale=.7]{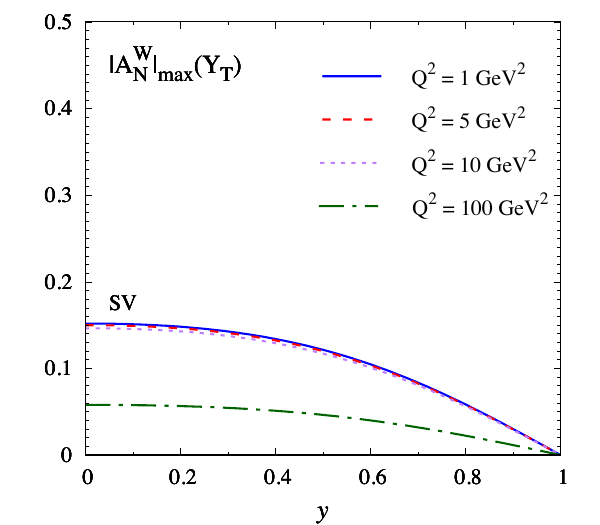}
\par\end{centering}
\caption{Same as in Figure~\ref{fig:cos2phiLsat}, but for transversely polarized $J/\psi$ (left panel) and $\Upsilon$ (right panel) mesons.
\label{fig:T} }
\end{figure}

\begin{figure}[t]
\begin{centering}
\includegraphics[clip,scale=.7]{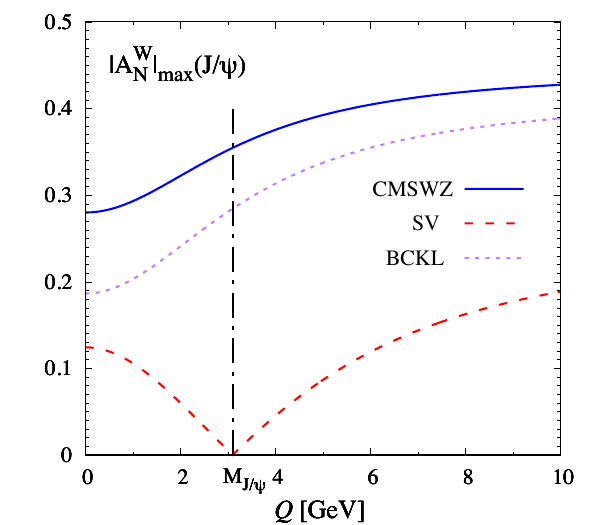}\includegraphics[clip,scale=.7]{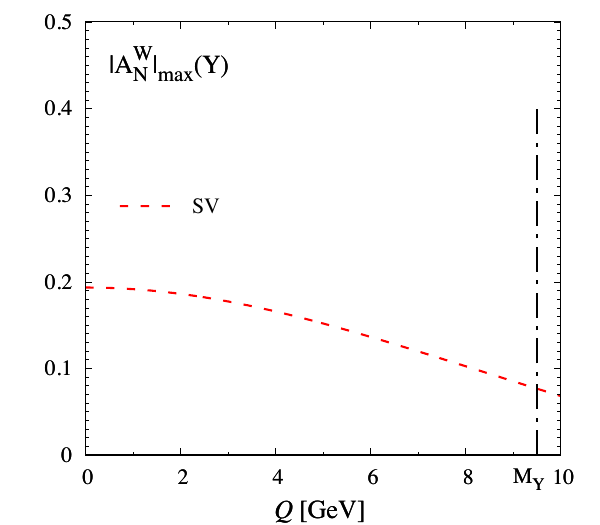}
\par\end{centering}
\caption{Upper bounds for the $\langle \cos 2\phi_\sT\rangle$ and  $A_N^W$  asymmetries, with $W =  \sin (\phi_S +\phi_\sT),\, \sin (\phi_S - 3 \phi_\sT)$, for $e \, p \to e^\prime \, J/\psi \, X $ (left panel) and $e \, p \to e^\prime \, \Upsilon\, X $ (right panel),  as a function of $Q$ for $y=0.1$. The labels  CMSWZ, SV and BCKL denote the adopted LDME sets, given in Tables \ref{tab:jpsiLDME} and \ref{tab:uLDME}.}
\label{fig:satQ} 
\end{figure}

First, in Fig.~\ref{fig:cos2phisat} we plot the maximum values of the azimuthal asymmetries $A_N^W$ computed for the processes 
$e \, p \to e^\prime \, J/\psi \, X $ (left panel) and $e \, p \to e^\prime \, \Upsilon \, X $ (right panel), in which the $J/\psi$ and $\Upsilon$ are unpolarized, as a function of $y$ and for different values of $Q^2$. The maximum values for $A_N^W$ in the case of longitudinally and transversely polarized quarkonium production are  presented in Figs.~\ref{fig:cos2phiLsat} and \ref{fig:T}, respectively.  In general, it turns out the asymmetries depend very strongly on the specific set of LDMEs adopted. As mentioned above, they always vanish in the limit $y\to1$, and reach their maximum when $y\to 0$. Alternatively,  in Fig.~\ref{fig:satQ} we show the $Q$-dependence of these maxima for the choice $y=0.1$ and only for unpolarized quarkonium production. The results for polarized quarkonia do not present significative differences and, for this reason, are not shown explicitly.


In the $J/\psi$ production plots one can see that depending on the set of LDMEs used and the quarkonium polarization state, the obtained behavior of the asymmetry as a function of $Q^2$ can be quite different, in some cases it increases or decreases, or even first decreases and then increases again. The $Q^2$ behavior of the asymmetries can thus be a further tool to determine the CO LDMEs, or at least their relative magnitude.

Finally, we point out that more stringent bounds  on gluon TMDs than the ones presented  in Eqs.~\eqref{eq:bounds}  can be obtained by comparison with experiments.  The COMPASS Collaboration has reported preliminary results on the Sivers asymmetry for $e\, p \, \to\, e^\prime\, J/\psi \, X$~\cite{Matousek:2016xbl}. The obtained value $A^{\sin(\phi_S -\phi_T)} = -0.28 \pm 0.18$ in the kinematical region of validity of our results, $z \ge 0.95$, points towards a negative gluon Sivers function at low $x$ and $Q^2$, with a size about $1/3$ of its positivity bounds.  An even smaller gluon Sivers function has been found  from the analyses of available data  on inclusive pion and heavy-quark production in proton-proton collisions at RHIC~\cite{DAlesio:2015fwo,DAlesio:2017rzj}. However, in those analyses, TMD factorization has been assumed for single-scale 
 processes, even if not supported by a formal proof. This phenomenological approach is known as generalized parton model (GPM) and is able to successfully describe many features of several available data.  It still remains to be seen whether the {\it effective} TMDs determined within the GPM differ from the ones extracted from TMD-factorizing processes.

\subsection{$\cos 2\phi $ asymmetries in the MV model}

In the small-$x$ limit, the nonperturbative McLerran-Venugopalan (MV) model \cite{McLerran:1993ni,McLerran:1993ka,McLerran:1994vd} allows to calculate the gluon distributions inside an unpolarized large nucleus or energetic proton. The analytical expressions for the unpolarized and linearly polarized Weizs\"acker-Williams (WW) gluon distributions in this model are given by \cite{Metz:2011wb,Dominguez:2011br}:
\begin{align}
f_1^g(x,\bm{p}_\sT^2)=\frac{S_\perp C_F}{\alpha_s \pi^3} \int \mathrm{d}r \frac{J_0(p_\sT r)}{r}\left (1-e^{-\frac{r^2}{4}Q_{sg}^2(r) }\right )\,,\\
h_1^{\perp g}(x,\bm{p}_\sT^2)=\frac{S_\perp C_F}{\alpha_s \pi^3} \frac{2 M_p^2}{p_\sT^2} \int \mathrm{d}r \frac{J_2(p_\sT r)}{r \ln \frac{1}{r^2 \Lambda^2}}\left (1-e^{-\frac{r^2}{4}Q_{sg}^2(r) }\right )\,,
\end{align}
where $S_\perp$ is the transverse size of the nucleus or proton, $\Lambda$ is an infrared cutoff such as $\Lambda_\mathrm{QCD}$, and where $Q_{sg}(r)$ is the saturation scale for gluons, which in the MV model depends logarithmically on the dipole size $r$, and in general is a function of $x$. Factorizing the dependence on $r$ as follows:  $Q\bm_{sg}^2(r)=Q_{sg0}^2 \ln{(1/r^2 \Lambda^2)} $, we can take $Q_{sg0}^2=(N_c/C_F) Q_{s0}^2$ with $Q_{s0}^2=0.35\,\mathrm{GeV}^2$ at $x=x_0=10^{-2}$ from the fits to HERA data \cite{GolecBiernat:1998js}. 

\begin{figure}[t]
\begin{centering}
\includegraphics[clip,scale=0.7]{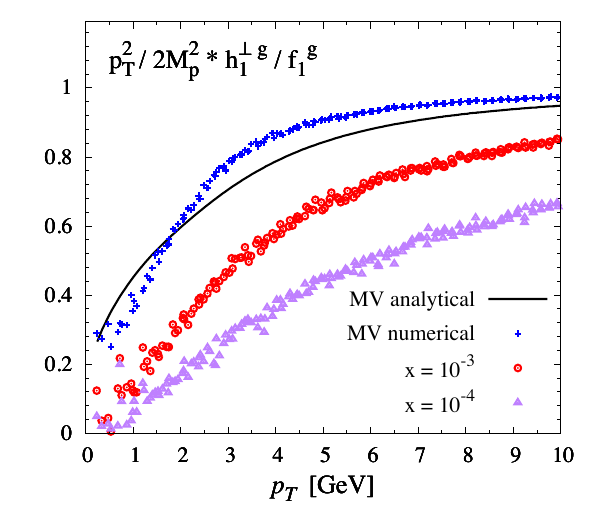}\includegraphics[clip,scale=0.7]{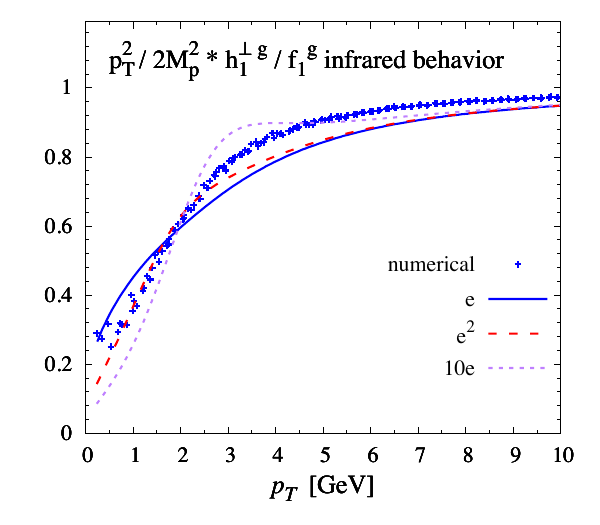}

\par\end{centering}
\caption{Left: ratio of the linearly polarized vs unpolarized Weizs\"acker-Williams gluon TMDs in the analytical MV model (solid line), as well as in the numerical simulation of Ref.~\cite{Marquet:2017xwy}. Right: the dependence of this ratio on the choice of IR regulator.}
\label{fig:boundMV} 
\end{figure}

Adding $e$ as a regulator for numerical convergence, in accordance with Ref.~\cite{Boer:2016fqd}, we obtain for the ratio of both gluon TMDs:
\begin{align}
\frac{p_{\sT}^{2}}{2M_{p}^{2}}\frac{h_{1}^{\perp g}(x,\bm{p}_{\sT}^{2})}{f_{1}^{g}(x,\bm{p}_{\sT}^{2})}=\frac{\int\mathrm{d}r\frac{J_{2}(p_{\sT}r)}{r\ln(\frac{1}{r^{2}\Lambda^{2}}+e)}\left (1-e^{-\frac{r^{2}}{4}Q_{sg0}^{2}\ln \left (\frac{1}{r^{2}\Lambda^{2}}+e \right )}\right )}{\int\mathrm{d}r\frac{J_{0}(p_{\sT}r)}{r}\left (1-e^{-\frac{r^{2}}{4}Q_{sg0}^{2}\ln \left (\frac{1}{r^{2}\Lambda^{2}}+e \right )}\right )}\,,
\label{eq:ratioanalytic} 
\end{align}
which is shown as a solid black line in the left panel of Fig. \ref{fig:boundMV}.


The nonlinear evolution in rapidity of the gluon density in the presence of saturation is governed by the JIMWLK equation \cite{Gelis:2010nm}, which can be solved numerically 
\cite{Rummukainen:2003ns,Lappi:2007ku,Lappi:2012vw}. In Refs.~\cite{Marquet:2016cgx,Marquet:2017xwy}, an implementation of JIMWLK on a two-dimensional lattice with spacing $a$ was used to evolve the gluon TMDs $f_1^g$ and $h_1^{\perp g}$ towards larger rapidities (or lower values of $x$), starting from an initial condition which corresponds to the MV model, using the prescription of Ref.~\cite{Lappi:2007ku}. We assigned a physical value to the lattice spacing $a$ according to the relation $a Q_{s0} =0.015$, which was obtained in Ref.~\cite{Marquet:2016cgx} by studying the universal large-$p_\sT$ behavior of the gluon TMDs. For our choice of the saturation scale, this relation yields $a=0.025\,\mathrm{GeV}^{-1} $. 
The numerical JIMWLK evolution is performed in steps $\delta s = (\alpha_s / \pi^2)\delta y=10^{-4}$, with $y=\ln{(x_0/x)}$ the rapidity. We show in Fig.~\ref{fig:boundMV} the initial condition at $y=0$ for the ratio $p_{\sT}^{2}/(2M_{p}^{2}) \,h_{1}^{\perp g}(x,\bm{p}_{\sT}^{2})/f_{1}^{g}(x,\bm{p}_{\sT}^{2})$, as well as the result after $500$ and $1000$ steps in the evolution, which at a coupling $\alpha_s(M_{J/\psi})\simeq 0.2$ corresponds to the values $x\simeq 10^{-3}$ and $x\simeq 10^{-4}$, respectively. 
\begin{figure}[t]
\begin{centering}
\includegraphics[clip,scale=0.7]{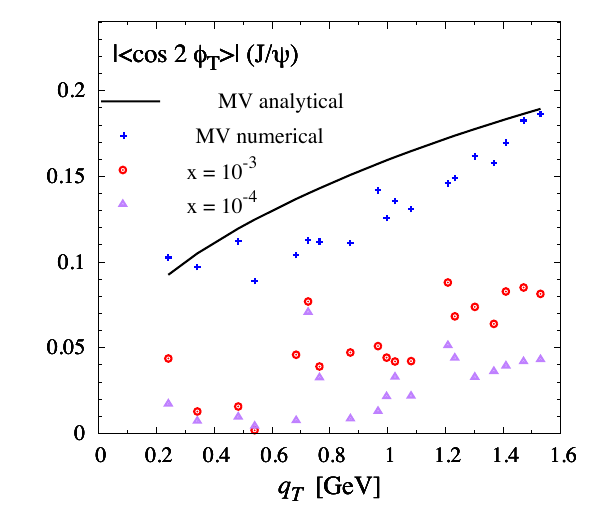}\includegraphics[clip,scale=0.7]{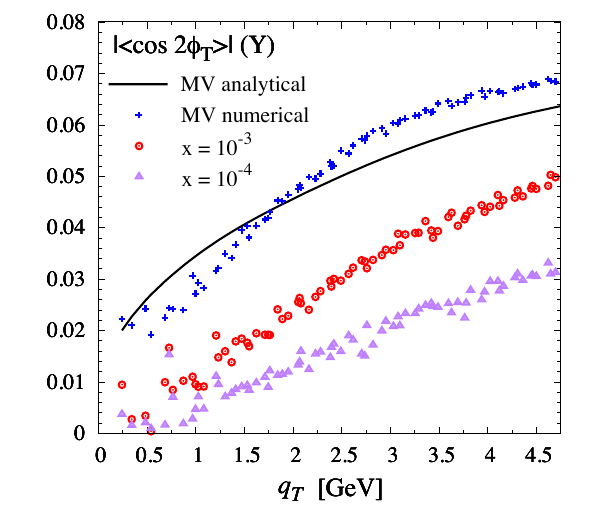}
\par\end{centering}
\caption{$\langle \cos 2\phi_T \rangle$ asymmetries as a  function of $q_T$  for the processes 
$e \, p \to e^\prime \, J/\psi \, X $ (left panel) and $e \, p \to e^\prime \, \Upsilon \, X $ (right panel), calculated at $Q=M_{\mathcal{Q}}$ and $y = 0.1$, and using the CMSWZ set for the $J/\psi$ LDMEs given in Table~\ref{tab:jpsiLDME}. }
\label{fig:cos2phivsq} 
\end{figure}

\begin{figure}[t]
\begin{centering}
\includegraphics[clip,scale=0.7]{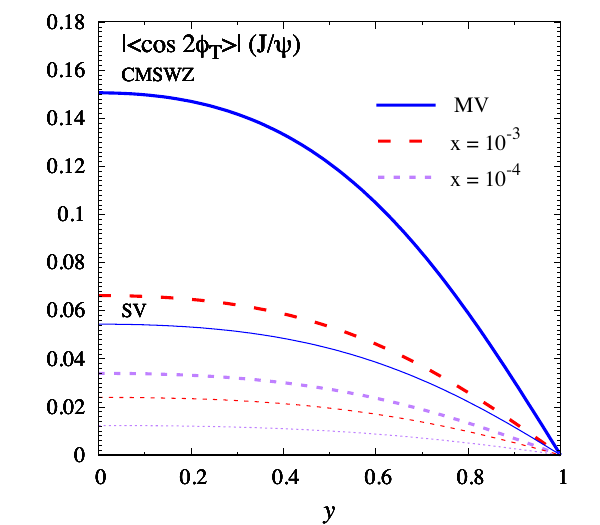}\includegraphics[clip,scale=0.7]{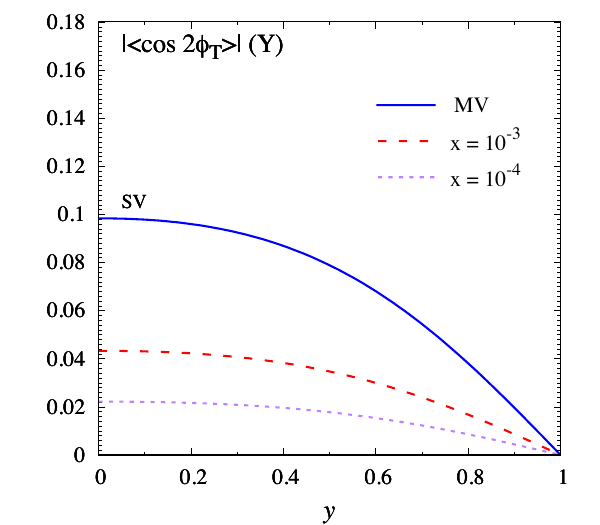}
\par\end{centering}
\caption{$\langle \cos 2\phi_\sT \rangle$ asymmetries as a  function of $y$  for the processes 
$e \, p \to e^\prime \, J/\psi \, X $ (left panel) and $e \, p \to e^\prime \, \Upsilon \, X $ (right panel), calculated at $Q = 1$  GeV and $q_\sT = 1.5\;\mathrm{GeV}$.  The labels  SV and CMSWZ denote the adopted LDME sets, given in Tables \ref{tab:jpsiLDME} and \ref{tab:uLDME}.}
\label{fig:cos2phivsy} 
\end{figure}

In the same Fig.~\ref{fig:boundMV}, we also compare with the analytical expression of the initial condition in Eq.~(\ref{eq:ratioanalytic}). It can clearly be seen  that the latter differs somewhat from its implementation on the lattice. Indeed, the uncertainty in the choice of the saturation scale and lattice size aside, the major source of this discrepancy is the way in which the infrared (IR) is regulated. On the lattice, this is taken care of by the finite lattice spacing, while in Eq.~(\ref{eq:ratioanalytic}) we added a regulator by hand. To illustrate the freedom in the way the IR can be regulated, and the ensuing theoretical uncertainty for the ratio of the TMDs, we plot this ratio in the right panel of Fig.~\ref{fig:boundMV} for some different choices of the regulator. 

With this ratio at hand, we show predictions for the $\cos{2\phi_\sT}$ asymmetry, Eq.~(\ref{eq:A0}), as a function of $q_\sT$ in Fig.~\ref{fig:cos2phivsq} and as a function of $y$ in Fig.~\ref{fig:cos2phivsy}, both for the $J/\psi$ and for the $\Upsilon$ mesons. Note that in Fig. \ref{fig:cos2phivsy} we took the numerical value for $p_{\sT}^{2}/(2M_{p}^{2})\,h_{1}^{\perp g}(x,\bm{p}_{\sT}^{2})/f_{1}^{g}(x,\bm{p}_{\sT}^{2})$ corresponding to $p_\sT=1.5\,\mathrm{GeV}$ from the numerical results in Fig.~\ref{fig:boundMV}, and then plotted the analytical $y$-dependence with the help of Eq.~(\ref{eq:A0}).  We restrict the range in $p_\sT$ to the region of validity of TMD factorization, which we estimate as $p_\sT = q_\sT\in[0,M_\mathcal{Q}/2]$. Note that in Fig.~\ref{fig:cos2phivsq}, the effects of the lattice discretization, which become more important for small values of $q_\sT$, are clearly visible, even suggesting a rise of the ratio towards $q_\sT\to 0$, which is clearly an artifact of the discretization.

\section{Summary and Conclusions}
\label{sec:conc}
We have presented expressions for the azimuthal asymmetries for $J/\psi$ and $\Upsilon$ production in DIS processes at LO in NRQCD, when the quarkonium in the final state is produced with a transverse momentum smaller than its invariant mass. Such observables can be used to extract information on the so far poorly known gluon TMDs, as well as to better understand the mechanism underlying quarkonium production by directly probing the two color-octet matrix elements  $\langle0\vert{\cal O}_{8}^{J/\psi}(^{1}S_{0})\vert0\rangle$ and $\langle0\vert{\cal O}_{8}^{J/\psi}(^{3}P_{0})\vert0\rangle$. We proposed ratios of asymmetries (Eqs.~(\ref{eq:ratioA1})-(\ref{eq:ratioA3})) in which the LDMEs cancel out, but also ratios (Eqs.~(\ref{eq:R2phi})-(\ref{eq:R}) and their polarized quarkonium analogues) in which the TMDs cancel out. The latter offer novel ways to extract the two mentioned CO LDMEs, which are still poorly known. This method requires one to consider comparisons of the processes $e\, p \to e^\prime \, {\cal Q}\, X$ and $e\, p \to e^\prime \,Q \,\overline{Q}\, X$ at the same hard scale, in order to establish a cancellation of the TMDs and to avoid having to include TMD evolution, although that in principle can be done with known methods as well.
Since the proposed asymmetries are always given by the ratio of two cross sections, they have the further advantage of being less sensitive to the normalizations of the cross sections, to the effects of higher order corrections and to other sources of uncertainties, like the exact value of the charm and bottom mass.  Moreover, as discussed in Ref.~\cite{Fleming:1997fq} to which we refer for details, we mention that for inclusive quarkonium production in DIS, nonperturbative effects like higher twist corrections and diffractive background can be more effectively suppressed as compared to quarkonium photoproduction, by looking at events with sufficiently high values of the photon virtuality $Q^2$. We have argued that in the kinematic region of high $Q^2$ and low transverse momenta considered, TMD factorization is applicable\footnote{For recent studies of the large $p_\sT$-range probed in $J/\psi + {\rm jet}$ production, see Refs.~\cite{DAlesio:2019qpk,Kishore:2019fzb}.} and the NLO CS contributions are suppressed. The latter can be suppressed further to a negligible level by an additional high lower-cut on $z$. This should ensure the robustness of the presented results. Final state transverse momentum smearing due to the transition from the CO $Q \overline{Q} $ state into the true CS hadronic final state is expected to affect the simplicity of the presented leading order results. Therefore, we checked its effect numerically using model parameterizations. From this we conclude that the approximation of ignoring $L$- and $M_{{cal Q}}$-dependent final state smearing effects may not be very important, especially if $\bm q_\sT^2$ is not too small. In addition, we have suggested ways to cross check the results between unpolarized and polarized quarkonium production.  As a by-product it offers a way to learn about the transverse momentum distribution of the hadronization into quarkonia, which is interesting in itself.

In our phenomenological analysis, we have used the positivity bounds for the gluon TMDs in order to estimate the maximal values of the asymmetries, and identify the kinematical regions in which they are sizable and where they could possibly be measured in future experiments at an EIC. For the gluon Sivers function this kind of studies provides little guidance because the corresponding maximal asymmetry is always 100\%. A more stringent constraint comes from preliminary COMPASS results which seem to suggest a nonzero Sivers effect. An investigation by the COMPASS Collaboration of the other asymmetries as well, in a kinematical region complementary to that of the EIC, would be of course very beneficial. Finding nonzero single spin asymmetries in quarkonium production would be a sign of the CO mechanism, because in the CS mechanism they have to vanish due to the absence of any initial or final state interactions~\cite{Yuan:2008vn}.

In addition to the upper bounds of the gluon TMDs, we have studied also their behavior at small values of $x$, which is expected to be probed at an EIC if $Q^2$ is not high, but of the order of $Q^2 \lesssim 10$ GeV$^2$ (also the higher $\sqrt{s}$ the better). The TMDs accessible in this process correspond to the so-called WW distributions at small $x$. This means that the $T$-odd distributions $f_{1T}^{\perp\,g}$, $h_{1T}^g$ and $h_{1T}^{\perp\,g}$ are suppressed by a factor of $x$ with respect to the unpolarized gluon density and, as such, cannot be described by current saturation models. On the other hand, the $T$-even distribution of linearly polarized gluons inside an unpolarized proton, $h_{1}^{\perp\,g}$, is not suppressed. We have therefore used the MV saturation model to show that it can give rise to sizable $\cos2\phi_\sT$ modulations.   Notably, we have explicitly checked that these modulations are only slightly reduced by small-$x$ evolution effects and could be in principle measured even down to $x = 10^{-4}$. \\

\begin{figure}[t]
\begin{centering}
\includegraphics[trim={6cm 23cm 4cm 3cm},clip,scale=1]{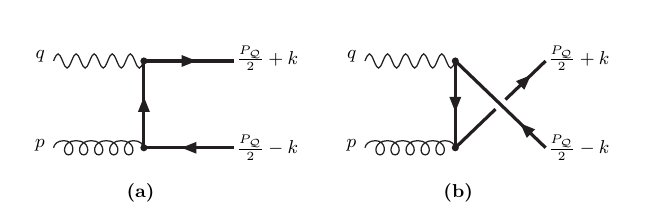}
\par\end{centering}
\caption{Feynman diagrams for the partonic subprocess $\gamma^* (q) \,+ \, g(p)\to {\cal Q} (P_{\cal Q})$ at LO in perturbative QCD.}
\label{fig:LO-ampl}
\end{figure}

\appendix
\section{Transition amplitudes}
\label{sec:app}

The scattering amplitudes for the partonic processes  $\gamma^*(q ) + g(p ) \to Q \overline Q  [^1S_0^{(8)}\, {\text{or}} \, \,^3S_1^{(8)} ] (P_{\cal Q} ) $  contributing to $ e\,p \to e^\prime \,J/\psi \,(\Upsilon)\,X$ can be written in the form~\cite{Ko:1996xw}
\begin{align}
{\cal{A}^{\mu\nu}} [^{1}S^{(8)}_{0} ] (q,p) &  =\frac{1}{4} \sqrt{\frac{ {\cal C}_{S,0}}{2 M_{\cal Q} }}\, {\rm Tr}\left [O^{\mu\nu}(0)\, (\Pqs - M_{\cal Q}) \,\gamma^5 \right ]\,,\label{eq:1S0} \\
{\cal{A}^{\mu\nu}} [^{3}S^{(8)}_{1} ] (q,p) &  =  \frac{1}{4} \sqrt{ \frac{ {\cal C}_{S,1}}{2 M_{\cal Q} }}\, {\rm Tr} \left [O^{\mu\nu}(0)\, (\Pqs - M_{\cal Q}) \,\epss_{S_z} \right ] \,, \label{eq:3S1}
\end{align} 
where
\begin{equation}
{\cal C}_{S,J} =  \frac{1}{2J+1}\,{ \langle 0 \vert {\cal O}_8^{J/\psi} (^{2 S+1} L_J)\vert 0 \rangle}\,,
\end{equation}
$M_{\cal Q} \approx 2 M_Q$ is the mass of the quarkonium, and  $\varepsilon_{S_z} $  is the  polarization vector for the spin-1 quarkonium wave function.  The operator ${O}(q,p; k, P_{\cal Q})$ is calculated at leading order in perturbative QCD from the Feynman diagrams in Fig.~\ref{fig:LO-ampl}, with $k$ being half the relative momentum of the outgoing quark-antiquark pair,  including the $SU(3)$ color-octet projector
\begin{align}
\langle  3 i , \bar {3} j\vert 8 a \rangle = \sqrt{2} \,  t^a_{ij}\, .
 \end{align}
We obtain
\begin{align}
O^{\mu\nu}(0) & =  -\sqrt{2}\, \delta^{ab} \frac{e e_c g_s}{2(M_{\cal Q}^2+Q^2)}\, \left [ \gamma^\mu \left ( \ps - \qs + M_{\cal Q} \right )\gamma^\nu \  - \gamma^\nu \left ( \ps -\,\qs - M_{\cal Q} \right )\gamma^\mu \right ]  \, ,
\end{align}
where we have used the shorthand notation ${O} (0) \equiv {O}(q,p; 0, P_{\cal Q})$. 

Similarly, the amplitudes for the subprocesses  $\gamma^*(q ) + g(p ) \to Q \overline Q  [^1P_0^{(8)}\, {\text{or}} \, \,^3P_J^{(8)} ] (P_{\cal Q} ) $, in which a $P$-wave bound state is formed, read 
\begin{align}
{\cal{A}^{\mu\nu}} [^{1}P^{(8)}_{1} ] (q, p)  & =  -i \sqrt{\frac{3\,{\cal C}_{P,0} }{8\,N_c\, M_{\cal Q}}} \, 
{\rm Tr}\left [ \left ({O^{\mu\nu}}(0)\,\epss_{L_z}\frac{\Pqs}{M_{\cal Q}} +   \varepsilon^{\alpha}_{L_z} \hat{O}^{\mu\nu}_{\alpha}(0)\, \frac{\Pqs-M_{\cal Q}}{2} \right ) \gamma^5 \right ]\,,\label{eq:1P0}\\
{\cal{A}^{\mu\nu}} [^{3}P^{(8)}_{0} ] (q,p) &  =   \frac{1}{2}\,i\,\sqrt{ \frac{{\cal C}_{P,0}}{{2\, N_c \,M_{\cal Q}}}}\,{\rm Tr}\left [ -3\,O^{\mu\nu}(0) + \left (\gamma^\alpha \hat{O}_{\alpha}^{\mu\nu}(0) -\frac{\Pqs P_{\cal Q}^\alpha}{M_{\cal Q}^2}\, \hat{O}_{\alpha}^{\mu\nu}(0) \right ) \frac{\Pqs - M_{\cal Q}}{2} \right ]\,, \label{eq:3P0} \\ 
{\cal{A}^{\mu\nu}} [^{3}P^{(8)}_{1}] (q,p) &  =  \frac{1}{4} \sqrt{\frac{3{\cal C}_{P,1}}{N_c\,M_{\cal Q}}} \, 
\epsilon_{\alpha\beta\rho\sigma}\,\frac{P_{\cal Q}^\rho}{M_{\cal Q}}\,\varepsilon_{J_z}^\sigma(P_{Q})\, {\rm Tr} \left [\gamma^\alpha \hat{O}^{\beta\mu\nu}(0) \,\frac{\Pqs - M_{\cal Q}}{2}   + O^{\mu\nu}(0)\, \frac{\Pqs}{M_{\cal Q}}\,\gamma^\alpha\gamma^\beta \right ]
\label{eq:3P1}   \,,\\
{\cal{A}^{\mu\nu}} [^{3}P^{(8)}_{2}] (q,p) & =  i\sqrt{\frac{3\,{\cal C}_{P,2}}{8\,N_c\,M_{\cal Q}}}\, 
\varepsilon^{\alpha\beta}_{J_z}(P_{\cal Q})\, {\rm Tr}\left [ \gamma_\alpha\,{\hat O}_\beta^{\mu\nu}(0)\,
\frac{\Pqs - M_{\cal Q}}{2} \right ]\,,
\label{eq:3P2}
\end{align}
where we have defined
\begin{equation}
{ \hat {O}_\alpha}^{\mu\nu}(0) \equiv \, \left . \frac{\partial }{\partial k^{\alpha}}{O}^{\mu\nu}(q, p; k)\right  \vert_{ k=0}\,,
\end{equation}
which can be calculated from the diagrams in Fig.~\ref{fig:LO-ampl} as well. Its explicit expression is given by 
\begin{align}
\hat{O}^{\alpha\mu\nu}(0) & = \sqrt{2}\, \delta^{ab} \frac{e e_c g_s}{M_{\cal Q}^2+Q^2}\,\, \left \{  \frac{2 p^ \alpha}{M_{\cal Q}^2+Q^2}   \left [ \gamma^\mu \left ( \qs - \ps - M_{\cal Q} \right )\gamma^\nu \  + \gamma^\nu \left ( \qs -\ps + M_{\cal Q} \right )\gamma^\mu \right ] - \gamma^\mu\gamma^\alpha\gamma^\nu - \gamma^\nu\gamma^\alpha\gamma^\mu \right \}\,. 
\end{align}

By computing the traces explicitly we obtain the following final results for the amplitudes:
\begin{align}
{\cal{A}}^{\mu\nu} [^{1}S^{(8)}_{0} ] (q,p) &  =  -2i\,   \sqrt{\frac{ {\cal C}_{S,0}}{M_{\cal Q}}} \, \delta^{a b}\, \frac{ee_cg_s}{M_{\cal Q}^2 +Q^2}\, {\epsilon^{\mu\nu}}_{\rho\sigma}\, q^{\rho} P_{{\cal Q }}^\sigma\,, \\
{\cal{A}}^{\mu\nu} [^{3}S^{(8)}_{1} ] (q,p) & = 0\,, \\
{\cal{A}}^{\mu\nu} [^{1}P^{(8)}_{1} ] (q,p) &  = 0\,, \\
{\cal{A}}^{\mu\nu} [^{3}P^{(8)}_{0} ] (q,p) &  =   {2 i}\, \sqrt{\frac{{\cal C}_{P,0}}{{N_c\, M_{\cal Q}}}}\, \delta^{a b}\,  \frac{ee_cg_s}{M_{\cal Q}^2 +Q^2}\,\frac{3M_{\cal Q}^2 +Q^2}{M_{\cal Q}}\,  \left [  g^{\mu\nu} - \frac{2}{M_{\cal Q}^2+Q^2}\, P_{\cal Q}^\mu\, q^{\nu}\right ]\, ,\\
{\cal{A}}^{\mu\nu} [^{3}P^{(8)}_{1} ] (q,p) & =  -4 \sqrt{\frac{9\,{\cal C}_{P,1}}{2\, N_c\,M_{\cal Q}}} \,\delta^{ab} \,\frac{ee_cg_s}{ (M_{\cal Q}^2 +Q^2)^2}\,  P_{\cal Q}^\rho \, \varepsilon_{J_z}^{\sigma} (P_{\cal Q})  \, \frac{Q^2}{M_{\cal Q}^2}\,\bigg [ (M_{\cal Q}^2+Q^2)\, {\epsilon_{\rho\sigma}}^{\mu\nu}  \nonumber \\
&\qquad \qquad \qquad \qquad \qquad +2\,  \epsilon_{\rho\sigma\alpha\beta}\, q^\alpha \left  (P_{\cal Q }^\mu\,  g^{\beta \nu} - P_{\cal Q}^\nu\,  g^{\beta \mu} \right )\bigg ]\, ,\\
{\cal{A}}^{\mu\nu} [^{3}P^{(8)}_{2} ] (q,p) &  =  2i\, \sqrt{ \frac{3\,M_{\cal Q}\, {\cal C}_{P,2}} {N_c} }\,\,\delta^{ab} \frac{ee_cg_s}{M_{\cal Q}^2 +Q^2}\,   \varepsilon_{J_z}^{\rho\sigma}(P_{\cal Q})
\bigg [ g^{\mu}_\rho g^{\nu}_\sigma + g^{\nu}_\rho g^{\mu}_\sigma \nonumber \\
& \qquad \qquad \qquad\qquad\qquad \qquad \qquad  - \frac{4}{M_{\cal Q}^2 + Q^2}\, q_{\sigma}  \left  ( g^{\mu\nu} q_{\rho} + g_{\rho}^{\nu } P_{{\cal Q}}^ \mu  -  g_{\rho}^{\mu } P_{{\cal Q}}^{\nu} \right)  \bigg ]\,.
\end{align}

Furthermore, in the calculation of the squared amplitudes we adopt some useful relations for the polarization vectors~\cite{Kuhn:1979bb,Guberina:1980dc}, which are reported below for completeness. If we denote by $\varepsilon^{\nu}_{J_z}$ the polarization vector for a bound state with total angular momentum $J=1$, four-momentum $P_{\cal Q}$ and mass $M_{\cal Q}$, then
\begin{equation}
\varepsilon_{J_z}^\alpha(P_{\cal Q}) \,P_{{\cal Q}\,\alpha} = 0\,, \qquad  \sum_{J_z = -1}^1 \varepsilon^\alpha_{J_z}(P_{\cal Q})\, \varepsilon^{*\beta}_{J_z}(P_{\cal Q}) = -g^{\alpha\beta}+\frac{P_{\cal Q}^\alpha P_{\cal Q}^\beta}{M_{\cal Q}^2} \equiv {\cal P}^{\alpha\beta}\,;
\label{eq:eps}
\end{equation}
while, if $\varepsilon^{\alpha\beta}_{J_z}$ is the polarization tensor for a $J =2$ system, then
\begin{eqnarray}
&&\qquad \varepsilon_{J_z}^{\alpha\beta}(P_{\cal Q}) = \varepsilon_{J_z}^{\beta\alpha}(P_{\cal Q}) \,, 
\qquad {\varepsilon_{J_z}^{\alpha}}_\alpha (P_{\cal Q}) = 0\, , \qquad  P_{{\cal Q}\,\alpha}\, \varepsilon_{J_z}^{\alpha\beta}(P_{\cal Q}) =0\,, \nonumber \\ 
&&  \sum_{J_z = -2}^2\varepsilon_{J_z}^{\mu\nu}(P_{\cal Q})  \,\varepsilon_{J_z}^{*\alpha\beta}(P_{\cal Q})   =  
\frac{1}{2}\, \left [ {\cal P}^{\mu\alpha} {\cal P}^{\nu\beta} + {\cal P}^{\mu\beta}{\cal P}^{\nu\alpha} \right ] - \frac{1}{3}\, {\cal P}^{\mu\nu} {\cal P}^{\alpha\beta}\,.
\label{eq:eps2}
\end{eqnarray}

Concerning the calculation of the cross section for the production of longitudinally and transversely polarized  spin-1 states,  
we notice that, if the $Q\overline Q$ pair is produced in a $^1S_0$ state, namely with $L=S=0$, the final quarkonium will be unpolarized. Therefore, for a $^1S_0^{(8)}$ configuration, each helicity state will contribute $1/3$ of the unpolarized cross section.  This explains the relative multiplicative factor of $\langle0\vert{\cal O}_{8}^{J/\psi}(^{1}S_{0})\vert0\rangle$ in Eq.~\eqref{eq:AUL_L}, which corresponds to a quarkonium that is longitudinally polarized, i.e.\ with helicity $\lambda=0$, with respect to the one in  Eq.~\eqref{eq:AUL}, which corresponds to an unpolarized quarkonium.  For $Q\overline Q$ pairs in $P$-wave intermediate states, with $L=S=1$, the method described above for the calculation of the unpolarized cross sections, consisting in the projection of the hard scattering amplitudes  onto states of definite quantum numbers $J$ and $J_z$, are not useful when the final quarkonium is polarized. Instead, one can project the amplitudes onto states of definite $L_z$ and $\lambda \equiv S_z$, square them and then sum over $L_z$.  The final results for the longitudinal and transversely polarized cross sections are obtained by using, respectively,  the following relations for the polarization vectors $\varepsilon_\lambda (P_{\cal Q})$ of the quarkonium~\cite{Budnev:1974de}, 
\begin{align}
 \varepsilon^\alpha_0(P_{\cal Q})\, \varepsilon^{*\beta}_{0}(P_{\cal Q})  & = \frac{P_{\cal Q}^\alpha P_{\cal Q}^\beta}{M_{\cal Q}^2} \, -\, \frac{P_{\cal Q}^\alpha\, n^\beta + P_{\cal Q}^\beta\, n^\alpha }{P_{\cal Q} \cdot n} \, + \,   \frac{M_{\cal Q}^2 \,n^\alpha\,n^\beta}{(P_{\cal Q} \cdot n)^2}   \,,
 \label{eq:hel-long}\\
\sum_{ \lambda  = \pm 1}  \varepsilon^\alpha_{\lambda}(P_{\cal Q})\, \varepsilon^{*\beta}_{\lambda}(P_{\cal Q})  & = -g^{\alpha\beta}  \,+\, \frac{P_{\cal Q}^\alpha\, n^\beta + P_{\cal Q}^\beta\, n^\alpha }{P_{\cal Q} \cdot n} \, - \,   \frac{M_{\cal Q}^2 \,n^\alpha\,n^\beta}{(P_{\cal Q} \cdot n)^2}   \,,
\label{eq:hel-transv}
\end{align}
where $n$ is any four-vector such that $n^2=0$ and $P\cdot n \neq 0$. Obviously, by summing Eqs.~\eqref{eq:hel-long} and \eqref{eq:hel-transv},  we obtain the second relation in Eq.~\eqref{eq:eps} with $J_z=\lambda$.
 
\acknowledgments
We would like to thank Jean-Philippe Lansberg, Cyrille Marquet, and Claude Roiesnel for useful discussions. This research is partially supported by the European Union's Horizon 2020 research and innovation programme (grant agreement No.~647981, 3DSPIN).

\end{document}